\documentclass[twocolumn,showpacs]{revtex4}
\usepackage[xdvi]{graphicx}
\usepackage{dcolumn}
\usepackage{amsmath}

\newcommand{\bra}{\left\langle}
\newcommand{\ket}{\right\rangle}
\newcommand{\pder}[2]{\frac{\partial #1}{\partial  #2}}
\newcommand{\pdert}[2]{\frac{\partial^2 #1}{\partial  #2^2}}
\newcommand{\der}[2]{\frac{d #1}{d  #2}}
\newcommand{\bv}[1]{{\boldsymbol #1}}
\newcommand{\e}{{\rm e}}

\newcommand{\ps}{p_{\rm s}}
\newcommand{\Js}{v_{\rm s}}
\newcommand{\rhod}{\rho_{\rm d}}
\begin{document}

\title{Systematic derivation of coarse-grained fluctuating hydrodynamic equations
for many Brownian particles under non-equilibrium condition}
\author{Takenobu Nakamura and Shin-ichi Sasa}

\date{\today}

\begin{abstract} 
We study the statistical properties of many Brownian particles 
under the influence of both a spatially homogeneous driving
force and a periodic potential with period $\ell$ in a 
two-dimensional space. In particular,
we focus on two asymptotic cases, $\ell_{\rm int} \ll \ell$ and 
$\ell_{\rm int} \gg \ell$, where $\ell_{\rm int}$ represents 
the interaction length between two particles. We derive 
fluctuating hydrodynamic equations describing the evolution 
of a coarse-grained density field 
defined on scales much larger than $\ell$ for both the cases.
Using the obtained equations, we calculate the equal-time 
correlation functions of the density field to the lowest order 
of the interaction strength. We find that the system exhibits
the long-range correlation of the type $r^{-d}$ ($d=2$) for the case
$\ell_{\rm int} \gg \ell$, while no such behavior
is observed for the case
$\ell_{\rm int}\ll \ell$.
\end{abstract}

\maketitle

\section{introduction}
 
To derive a description of macroscopic phenomena 
on the basis of a microscopic model 
is an important problem in statistical physics.
In equilibrium systems,
equilibrium statistical mechanics provides a clear solution to
this problem. 
However, when a system is out of 
equilibrium, even in a non-equilibrium steady state,
no general framework is known except for linear response theory 
that is applied to systems close to equilibrium \cite{Kubo}.
Here, let us recall 
Boltzmann's significant research that led to the genesis of equilibrium 
statistical mechanics. He arrived at his famous formula by analyzing
the simplest system, a dilute gas, thoroughly. Thus, in order to 
approach non-equilibrium statistical mechanics,
we investigate a simple non-equilibrium system with exploring
a relation between microscopic and macroscopic descriptions. 

Let us consider a system in which many sub-micrometer particles 
are driven by an external 
force in a solvent.
The force consists of both a spatially homogeneous
driving force and the force generated by a spatially periodic potential.
It has been known that such a system can exhibit phenomena out of
local equilibrium 
and that it can be designed for experiments 
due to the recent development of the optical technology used 
in controlling and measuring particles \cite{Grier,Grier2}.
The system also provides a typical example of the so-called 
{\it driven diffusive system} \cite{Schmidttmann,Gregory}.
Theoretically, there exists a case that the motion of the particles is accurately described 
by a Langevin equation.
In this system, we can investigate
macroscopic phenomena both by 
a laboratory experiment
and by an analysis 
on the basis of the Langevin equation that is regarded as 
a  microscopic model. 
In particular, the systematic calculation of 
macroscopic quantities from the microscopic model is
the first step in developing a new framework of non-equilibrium 
statistical mechanics.

Macroscopic phenomena in driven diffusive systems
have been described phenomenologically within a framework of 
fluctuating 
hydrodynamics \cite{Grinstein,Garrido,Gregory}.
The dynamical variable in this description is a density field,
and its evolution equation is assumed 
to take the simplest form under the imposed physical requirements.
For example, 
it is assumed that the evolution equation for
the density field in driven diffusive systems 
possesses an anisotropic nature and no detailed balance condition.
With this simple assumption, the existence of a long-range 
correlation in driven diffusive systems was predicted
even in a linear model \cite{Dorfman,Grinstein}.
Furthermore, the anomalous behavior 
of the space-time correlation function of density fluctuations has 
been studied by analyzing a non-linear model 
for driven diffusive systems \cite{BKS,NOS}.

It is expected that the macroscopic behavior in 
driven Brownian particle systems is described by a 
fluctuating hydrodynamic equation.
We then address a problem
to quantitatively derive the form of fluctuating hydrodynamic equations 
on the basis of a microscopic model describing the motion of the 
particles. If this problem is solved, 
we can calculate the 
correlation function of density fluctuations
by using the obtained fluctuating hydrodynamic equation.
Then, the calculation 
result is more quantitative than that by a fluctuating hydrodynamic equation
assumed phenomenologically.
In general, it is believed that 
the equal-time correlation function in $d$-dimensional 
driven diffusive systems exhibits the 
power-law behavior of the type $r^{-d}$ in the long distance regime 
and has a short-range part that deviates from the correlation 
determined by 
equilibrium statistical mechanics.
However, recently, 
it has been shown that this type of long-range correlation 
does not appear in some lattice gases \cite{Tasaki}.
Further, the short-range part of 
the correlation is connected with an extended thermodynamic function 
in driven lattice gases \cite{HS1,SST}.
Therefore, it is important to calculate a concrete form of the correlation function for driven Brownian particle systems.

Taking these into consideration, 
in this paper, we derive evolution 
equations of a coarse-grained 
density field from a many-body Langevin equation describing the motion 
of Brownian particles under a non-equilibrium condition.
We first note that the many-body Langevin 
equation is equivalent to a non-linear fluctuating equation for
the density field \cite{Dean}. 
To the latter equation, 
we apply a system reduction method
in order to describe the large-scale dynamics of the density field
\cite{Kuramoto,CrossHohenberg}.
As a result, we obtain the evolution equation of the coarse-grained 
density field, and we calculate the equal-time correlation function.
Furthermore, with regard to the calculation method,
we extend a standard system reduction method
so as to analyze 
a stochastic partial differential equation.
Therefore, we expect that this paper contributes to the progress 
of such perturbation theory, too.

This paper is organized as follows. In Sec. \ref{model},
we introduce a Langevin equation describing the motion of
many Brownian particles under an external force. We then 
obtain a non-linear fluctuating hydrodynamic equation 
for the density field. In Sec. \ref{Analysis},
we develop a perturbation method to derive a coarse-grained 
fluctuating hydrodynamic equation and calculate 
the equal-time correlation functions of density fluctuations. 
Section \ref{final} is devoted to
remarks on the present study. The technical details are 
presented in Appendixes.

\section{model}\label{model}

We study 
$N$ Brownian particles suspended in a two-dimensional solvent
of temperature $T$.
Let $\bv{x}_{i}$ be the position of the $i$-th particle. 
Here, $i$ is an integer satisfying $1 \le i \le N$.
We express the $\alpha$-th component of $\bv{x}_{i}$ as $x_{i\alpha}$, 
with $\alpha=1,2$. 
That is, $\bv{x}_{i}=(x_{i1},x_{i2})$. 
In addition, 
$\bv{x}=(x_1,x_2)$ indicates a position in the two dimensional 
space, where $0 \le x_{i\alpha} \le L$, 
and periodic boundary conditions are 
assumed for simplicity.
Each particle is driven by an external force $f \bv{\e}_1=(f,0)$ 
and is subject to a periodic potential $U(x_1)$ with period $\ell$
such that $L/\ell$ is an integer.
For simplicity, we assume that the periodic potential is independent of 
$x_2$.
Furthermore, we express the interaction between the $i$-th 
particle and the $j$-th particle by an interaction potential 
$u(\bv{x}_i-\bv{x}_j)$. 
We assume that the function $u(\bv{r})$ decays to zero with 
a typical length, $\ell_{\rm int}$.
$\bar{\rho}=N/L^2$ represents the average density 
of the Brownian particles.

The motion of the $i$-th Brownian particle is assumed to be described by
a Langevin equation
\begin{eqnarray}
\gamma\der{x_{i \alpha}}{t}&=&
\left(f-\pder{U(x_{i1})}{x_{i \alpha}}\right)\delta_{\alpha 1}\nonumber\\
&&-\pder{ \ }{x_{i \alpha}}
\sum_{j=1,j\neq i}^{N}u(\bv{x}_i-\bv{x}_j)
+R_{i \alpha}(t),
\label{lange}
\end{eqnarray}
where $\gamma$ is a friction constant and 
$R_{i \alpha}$ is zero-mean Gaussian white noise 
that satisfies
\begin{equation}
\bra R_{i \alpha}(t)R_{j \beta}(t')\ket 
=2\gamma T\delta_{\alpha\beta}\delta_{ij}\delta(t-t').
\label{noiseparticle}
\end{equation}

It should be noted that without the periodic potential $U$,
the system is equivalent to an equilibrium system in a 
moving frame with velocity $f/\gamma$.
Thus, the periodic potential is necessary 
for investigating the non-equilibrium nature.
Note that such a periodic potential 
can be implemented experimentally by using an optical 
phase modulator \cite{Grier,Grier2}.

Now, we define the fine-grained density field as
\begin{equation}
\rhod(\bv{x},t)\equiv\sum_{i=1}^N\delta^2(\bv{x}-\bv{x}_i(t)).
\label{dens}
\end{equation}
This satisfies the continuity equation
\begin{eqnarray}
\pder{\rhod(\bv{x},t)}{t}
=-\sum_{\alpha=1}^2 \pder{j_\alpha(\bv{x},t)}{x_\alpha},
\label{conti}
\end{eqnarray}
and the expression of current $j_\alpha(\bv{x},t)$ is derived from
Eq. (\ref{lange}) with Eq. (\ref{noiseparticle}) as follows:
\begin{eqnarray}
j_\alpha(\bv{x},t)&=&
\frac{\rhod(\bv{x},t)}{\gamma}
\left[
   f\delta_{\alpha 1}-\pder{ \ }{x_\alpha}
  \left(
     \left. \frac{\delta H}{\delta \varphi(\bv{x})} 
     \right\vert_{\varphi=\rhod(\ ,t)} 
  \right) 
\right]\nonumber\\
&&+\sqrt{\frac{T \rhod(\bv{x},t)}{\gamma}}\xi_\alpha(\bv{x},t),
\label{devdens}
\end{eqnarray}
where the real-valued functional $H$ for a function $\varphi(\bv{x})$ 
takes the form
\begin{eqnarray}
H(\varphi)&=&\int {\mathrm d}^2\bv{x}\varphi(\bv{x})U(\bv{x})\nonumber\\
&&+\frac{1}{2}\int {\mathrm d}^2\bv{x}\int {\mathrm d}^2\bv{y}
\varphi(\bv{x})u(\bv{x}-\bv{y})\varphi(\bv{y})\nonumber\\
&&+T\int {\mathrm d}^2\bv{x}\varphi(\bv{x})[\log\varphi(\bv{x})-1],
\label{Hamiltonian}
\end{eqnarray}
and $\xi_\alpha(\bv{x},t)$ in Eq. (\ref{devdens}) 
represents zero-mean space-time Gaussian noise satisfying
\begin{equation}
\bra \xi_\alpha(\bv{x},t) \xi_\beta(\bv{x}',t') \ket 
=2\delta_{\alpha \beta}\delta^2(\bv{x}-\bv{x}')\delta(t-t').
\label{noise}
\end{equation}
Throughout this paper, the multiplication of $\xi_\alpha(\bv{x},t)$
with a usual function $\phi(\bv{x},t)$ is 
interpreted as the Stratonovich rule in the space variable
and the Ito rule in the time variable.
The derivation of fluctuating hydrodynamic equations
for the fine-grained density
was reported in Ref. \cite{Dean}.
We explain a derivation method 
of Eq. (\ref{conti}) with Eqs. (\ref{devdens}), (\ref{Hamiltonian}), and 
(\ref{noise}) in Appendix \ref{deridens}.

In the equilibrium case when $f=0$, the form of Eq. (\ref{devdens}) is 
understood by the following physical considerations. The first 
term in Eq. (\ref{devdens}) represents
a drift caused by the gradient of the chemical potential that is 
given by the functional derivative of the potential $H(\rhod)$
with respect to $\rhod$. 
Here, the functional form of 
$H(\rhod)$ 
may be physically interpreted 
by noting that the third term in Eq. (\ref{Hamiltonian})
represents the entropy contribution  of non-interacting particles.

The noise term in Eq. (\ref{devdens}) can be understood from 
the fact that the system with $f=0$ satisfies the detailed balance 
condition. This is verified as follows. Let $\Delta t$ be a small 
time interval and ${\rm Tr}(\varphi \to \varphi')$ be 
the conditional probability 
of the density profile $\varphi'(\bv{x})$ at time $t+\Delta t$,  
provided the density  profile is $\varphi(\bv{x})$ at time $t$. 
(
${\rm Tr}(\varphi\rightarrow\varphi')$ is referred to as the transition probability
from $\varphi$ at time $t$ to $\varphi'$ at time $t+\Delta t$.)
For the case $f=0$,
using Eqs. (\ref{conti}), (\ref{devdens}), and (\ref{noise}),
we  calculate 
\begin{eqnarray}
{\rm Tr}(\varphi \to \varphi')\nonumber\\
=\frac{1}{{\cal N}}
\exp\Big[
&-&\Delta t \int {\mathrm d}^2\bv{x}
\frac{\varphi(\bv{x})}{4\gamma T}\left(\vec{\nabla}
\frac{\delta H(\varphi)}{\delta \varphi(\bv{x})}\right)^2
\nonumber \\
&-&\Delta t \int {\mathrm d}^2\bv{x}
\frac{\gamma}{4T\varphi(\bv{x})}
\left[\vec{\nabla} \Delta^{-1}
\left(\frac{\varphi'(\bv{x})-\varphi(\bv{x})}{\Delta t}\right)
\right]^2\nonumber\\
&-&\frac{\Delta t}{2T}\int {\mathrm d}^2\bv{x}
\frac{\varphi'(\bv{x})-\varphi(\bv{x})}{\Delta t}
\frac{\delta H(\varphi)}{\delta \varphi(\bv{x})}\nonumber\\
&+&O\left((\Delta t)^2\right)
\Big],
\label{trdens1}
\end{eqnarray}
where $\Delta^{-1}$ is the Green function of the Laplacian operator,
and ${\cal N}$ is the normalization constant determined by 
\begin{equation}
\int {\cal D} \varphi'  
{\rm Tr}(\varphi \to \varphi')=1,
\label{normal}
\end{equation}
where ${\cal D}\phi'$ represents a functional measure.
From Eq. (\ref{trdens1}), we obtain 
\begin{equation}
\frac{{\rm Tr}(\varphi \to \varphi')}
{{\rm Tr}(\varphi' \to \varphi)}
=\exp\left[-\frac{H(\varphi')-H(\varphi)}{T} 
+O\left((\Delta t)^2\right) \right]. 
\label{DB}
\end{equation}
Using Eqs. (\ref{normal}) and (\ref{DB}), we derive  
\begin{equation}
\int {\cal D} \varphi  P_{\rm S}(\varphi)
{\rm Tr}(\varphi \to \varphi')
=P_{\rm S}(\varphi'),
\label{steady}
\end{equation}
where
\begin{equation}
P_{\rm S}(\rhod)=\frac{1}{Z_{\rm f}}
\exp\left[-\frac{H(\rhod)}{T}\right].
\label{can}
\end{equation}
Eq. (\ref{steady}) implies that $P_{\rm S}(\rhod)$ is a steady distribution 
functional of the density profile $\rhod$. 
Then, Eq. (\ref{DB}) is rewritten as
\begin{align}
P{\rm s}(\varphi)&{\rm Tr}(\varphi \to \varphi')\nonumber\\
&=P{\rm s}(\varphi'){\rm Tr}(\varphi' \to \varphi)
\exp(O\left((\Delta t)^2\right) ).
\label{DB2}
\end{align}
This is the detailed balance condition with respect to the 
distribution given in Eq. (\ref{can}).
It should be noted that $\rhod$ in the third term 
in Eq. (\ref{devdens}) must not be replaced with its average
value, because this replacement breaks the detailed balance
condition.

In the non-equilibrium case,
the effect of the external force $f$ is expressed only 
by the first term in Eq. (\ref{devdens}) as a modification of the gradient
of the chemical potential.
Although the modification from the equilibrium case is minimum,
the modification breaks the detailed balance condition.
Therefore, we need to analyze the evolution equation
in order to obtain the steady probability distribution of the density field.

\section{Analysis}\label{Analysis}

This section consists of  four subsections. In Sec. \ref{basicframework}, we
present the basic framework of our perturbation method.
In Sec. \ref{no-interacting}, we investigate a non-interacting case
as the simplest example. 
We find that there is no long-range correlation of 
the density field for the non-interacting systems.
In Sec. \ref{weak-interacting}, 
we take into account the effects of 
the particle interaction perturbatively.
In particular, we focus on 
two asymptotic cases $\ell_{\rm int} \gg \ell$ and $\ell_{\rm int} \ll \ell$.
In Sec. \ref{correlation}, 
we calculate the correlation functions of the density 
field for the two cases and demonstrate 
that there exists the long-range correlation of
the type $1/r^d$ only for the case $\ell_{\rm int}\gg\ell$. 

\subsection{basic framework}\label{basicframework}

We analyze the non-linear fluctuating hydrodynamic equation given 
in Eq. (\ref{conti}) 
with Eqs. (\ref{devdens}), (\ref{Hamiltonian}), and (\ref{noise}).
Our perturbation method is based on the expansion of 
the weak interaction,
weak noise and the separation of length scales. In order to represent
the expansion parameters explicitly, we replace $\xi_\alpha(\bv{x},t)$ 
and $u(\bv{x}-\bv{y})$ with $\mu \xi_\alpha(\bv{x},t)$ and 
$\lambda u(\bv{x}-\bv{y})$, respectively. We also set 
$\epsilon \equiv \ell/L$. The parameters $\mu$, $\epsilon$ and  
$\lambda$ are regarded as  small parameters in our analysis.

With this setting, we first consider the case that $\mu=0$ and 
$\lambda=0$ in the equation to be analyzed. Because the evolution equation is
deterministic in this case, the density field relaxes to the steady 
one $\rhod^{(0)}$ that satisfies 
\begin{eqnarray}
\bar\rho\Js=-\frac{1}{\gamma}\left(\pder{U(x_1)}{x_1}-f+T\pder{ \ }{x_1}
\right)\rhod^{(0)}(\bv{x}),
\label{steadyrhod0}
\end{eqnarray}
where $\Js$ represents the average velocity of the particle 
in the steady state,
and  it is expressed as \cite{HS2,Reimann1,Reimann2}
\begin{eqnarray}
\Js=\frac{T}{\gamma}
(1-\e^{-\beta f\ell})
\left(\int_0^\ell \frac{{\mathrm d}x_1}{\ell}I_-(x_1)\right)^{-1}.
\label{defofJs}
\end{eqnarray}
Here, the function $I_\pm(x_1)$ is defined as
\begin{eqnarray}
I_\pm(x_1)=\int_0^\ell {\mathrm d}x_1'
\e^{\pm\beta U(x_1)\mp\beta U(x_1\mp x_1')-\beta fx_1'}.
\label{Ipm:def}
\end{eqnarray}

By using the condition 
\begin{equation}
\int_0^\ell \frac{{\mathrm d}x_1}{\ell} \rhod^{(0)}(\bv{x})=\bar \rho,
\end{equation}
we derive the solution of Eq. (\ref{steadyrhod0}) as follows: 
\begin{equation}
\rhod^{(0)}(\bv{x})=\ps(x_1)\bar \rho,
\label{zero}
\end{equation}
where 
\begin{eqnarray}
\ps(x_1)=\frac{1}{Z}I_-(x_1).
\end{eqnarray}
The normalization factor  $Z$  is determined by the condition
\begin{eqnarray}
\int_0^\ell {\mathrm d} x_1 \ps(x_1)= \ell.
\label{normalization}
\end{eqnarray}

Now, we consider the case 
that $\mu \not =0$ and $\lambda \not =0$. 
Based on Eq. (\ref{zero}), we set 
\begin{eqnarray}
\rhod(\bv{x},t)=\ps(x_1)q(\bv{x},t).
\label{defofq}
\end{eqnarray}
That is, the variable $q(\bv{x},t)$ is equal to the average density
$\bar \rho$ when $\mu  =0$ and $\lambda  =0$, and this variable 
represents
the density modulation caused by the noise and interaction. 
Substituting Eq. (\ref{defofq}) into Eqs. (\ref{conti}), (\ref{devdens}), 
and (\ref{Hamiltonian}) with the replacement explained in the first
paragraph in this subsection, we obtain the evolution equation for 
$q(\bv{x},t)$ as
\begin{eqnarray}
\pder{q(\bv{x},t)}{t}&=&\hat{M}q(\bv{x},t)\nonumber\\
&&-\mu\frac{1}{\ps(x_1)}\sqrt{\frac{T}{\gamma}}
\pder{ \ }{\bv{x}}\cdot\left(\sqrt{\ps(x_1)q(\bv{x},t)}
\bv{\xi}(\bv{x},t)\right)\nonumber\\
&&-\lambda\frac{1}{\ps(x_1)}
\pder{ \ }{\bv{x}}\cdot \bv{j}_{\rm int}(\bv{x},t),
\label{qdev}
\end{eqnarray}
where the operator $\hat M$ is calculated as 
\begin{eqnarray}
\hat{M}&=&\left(-\frac{\Js}{\ps(x_1)}
+\frac{T}{\gamma}\pder{\log \ps(x_1)}{x_1}
\right)\pder{ \ }{x_1}\nonumber\\
&&+\frac{T}{\gamma}\left(\pdert{ \ }{x_1}+\pdert{ \ }{x_2}\right),
\end{eqnarray}
and $\bv{j}_{\rm int}(\bv{x},t)$ is defined as
\begin{align}
&\bv{j}_{\rm int}(\bv{x},t)\nonumber\\
&=\frac{1}{\gamma}\ps(x_1)q(\bv{x},t)
\int {\mathrm d}^2\bv{y}\pder{u(\bv{x}-\bv{y})}{\bv{x}}\ps(y_1)q(\bv{y},t),
\label{jint}
\end{align}
which represents a current generated by the particle interaction.

Next, we notice the separation of length scales ($\epsilon  \ll 1$). 
We pay attention to the density fluctuations on length
scales of order $L$ and introduce the density field $Q(\bv{X},t)$ with
a large scale coordinate $\bv{X}=\epsilon \bv{x}$. 
While, on length scales of order $\ell$, 
the periodic potential determines the 
system behavior. In order to express this in an explicit manner, 
we introduce a phase variable $\theta$ as $\theta={\rm mod}(x_1,\ell)$
\cite{Kuramoto,CrossHohenberg}.
Obviously, $U(\theta)=U(x_1)$ and $\ps(\theta)=\ps(x_1)$.

The density fluctuations with a smaller wave-number have a longer time
scale. Then, when we focus on a time scale related to diffusion 
in the entire system, 
we assume that  $Q(\bv{X},t)$ obeys an autonomous equation
and that $q(\bv{x},t)-Q(\bv{X},t)$ depends on time $t$
only through the density field $Q(\bv{X},t)$. These assumptions are
expressed as
\begin{align}
\pder{Q}{t}&=\Omega(Q), 
\label{Qev-gen} \\
q(\bv{x},t) &= Q(\bv{X},t)+\rho(\theta,Q),
\label{s-mfd}
\end{align}
where $\Omega(Q)$ represents a map providing
a function of $(\bv{X},t)$
for the density field $Q$,
and $\rho(\theta,Q)$ represents a similar map for each $\theta$
(see Ref. \cite{Kuramoto,CrossHohenberg}). 

Hereinafter, we treat $\theta$ and $X$ as independent variables. 
Then, when the spatial derivative acts on 
a function of $(\theta,\bv{X})$, it should be read as 
\begin{eqnarray}
\pder{ \ }{\bv{x}}=\pder{ \ }{\theta}\bv{\e}_1+\epsilon\pder{ \ }{\bv{X}}.
\label{multi}
\end{eqnarray}
Furthermore, the two quantities 
$\bv{\xi}(\bv{x},t)$ and $u(\bv{x})$ in Eq. (\ref{qdev}) 
with Eq. (\ref{jint})
can be expressed as  $\bar{\bv\xi}(\theta,\bv{X},t)$ 
and $\bar u(\theta,\bv{X})$, respectively,
by using a method presented in Appendix \ref{treatmulti}.
Using this expression, Eq. (\ref{noise}) leads to
\begin{align}
&\bra\bar\xi_\alpha(\theta,\bv{X},t)
\bar\xi_\beta(\theta',\bv{X}',t')\ket\nonumber\\
&=2\delta_{\alpha\beta}
\ell\delta(\theta-\theta')\epsilon^2\delta^2(\bv{X}-\bv{X}')
\delta(t-t'),
\label{multi:noise}
\end{align}
as explained in Appendix \ref{treatmulti}.
Further, using $\bar u (\theta,\bv{X})$, $\bv{j}_{\rm int}(\bv{x},t)$
in Eq. (\ref{jint})  is expressed as
\begin{align}
\bar{\bv{j}}_{\rm int}&(\theta,\bv{X},t)\nonumber\\
=&\ps(\theta)\left[Q(\bv{X},t)+\rho(\theta,Q)\right]\nonumber\\
&\left(\pder{ \ }{\theta}\bv{\e}_1+\pder{ \ }{\bv{X}}\right)
\int \frac{{\mathrm d}\theta'}{\ell}
\int \frac{{\mathrm d}^2\bv{Y}}{\epsilon^2}
\bar u(\theta-\theta',\bv{X}-\bv{Y})\nonumber\\
&\ps(\theta')\left[Q(\bv{Y},t)+\rho(\theta',Q)\right],
\label{multi:jint}
\end{align}
where we have used Eq. (\ref{multi:integration}).
Finally, $\hat{M}$ can be expanded as
\begin{equation}
\hat{M}=\hat{M}^{(0)}+\epsilon\hat{M}^{(1)}+\epsilon^2\hat{M}^{(2)},
\label{M-exp}
\end{equation}
where $\hat{M}^{(i)}$ with $i=0,1$ and $2$ are calculated as
\begin{align}
\hat{M}^{(0)}=&\left(-\frac{\Js}{\ps(\theta)}
+\frac{T}{\gamma}\der{\log \ps(\theta)}{\theta}\right)
\pder{ \ }{\theta}
+\frac{T}{\gamma}\pdert{ \ }{\theta},
\label{m0}\\
\hat{M}^{(1)}=&\left(-\frac{\Js}{\ps(\theta)}
+\frac{T}{\gamma}\der{ \log \ps(\theta) }{\theta}
+2\frac{T}{\gamma}\pder{ \ }{\theta}\right)\pder{ \ }{X_1},
\label{m1}\\
\hat{M}^{(2)}=&\frac{T}{\gamma}
\left(\pdert{ \ }{X_1}+\pdert{ \ }{X_2}\right).
\label{m2}
\end{align}

Using these, the substitution of Eqs. (\ref{Qev-gen}), 
and (\ref{s-mfd})
into Eq. (\ref{qdev}) yields
\begin{eqnarray}
&&\Omega(Q)+\frac{\delta \rho}{\delta Q}\cdot \Omega(Q)\nonumber\\
&=&\hat{M}\left[Q+\rho(\theta,Q)\right]\nonumber \\
&&-\mu \frac{1}{\ps(\theta)}\sqrt{\frac{T}{\gamma}}
\left(\pder{ \ }{\theta}\bv{\e}_1+\epsilon\pder{ \ }{\bv{X}}\right)
\nonumber\\
&&\cdot \left[\sqrt{\ps(\theta)
(Q+\rho(\theta,Q))}\bar{\bv{\xi}}(\theta,\bv{X},t)\right]\nonumber \\
&&-\lambda\frac{1}{\ps(\theta)}
\left(\pder{ \ }{\theta}\bv{\e}_1+\epsilon\pder{ \ }{\bv{X}}\right)
\cdot \bar{\bv{j}}_{\rm int}(\theta,\bv{X},t),
\label{Q-rho}
\end{eqnarray}
where $(\delta \rho/\delta Q)$ is the operator on
a function $\varphi(\bv{x})$ defined as 
\begin{equation}
\rho(\theta, Q+\varphi)-\rho(\theta, Q)=
\frac{\delta\rho}{\delta Q}\cdot \varphi(\bv{x}) +O(|\varphi|^2),
\end{equation}
in the limit $|\varphi| \to 0$ with an appropriate norm $|\cdot|$ 
of the function space. 
Recall that $\rho$ is a map in the function space for each $\theta$.
Thus, 
the operator $(\delta \rho/\delta Q)$ mathematically 
corresponds to a Fr\'{e}chet derivative
when the function space is properly defined.
This should not be confused with the functional derivative used in 
Eq. (\ref{devdens}).

Now, we derive $\Omega(Q)$ and $\rho(\theta,Q)$  by solving 
Eq. (\ref{Q-rho}) with the perturbation method. 
Because we assumed that $\epsilon$, $\mu$ and $\lambda$ are small,
we expand $\rho(\theta,Q)$ and $\Omega(Q)$ in these parameters. 
Specifically, setting $\mu=\epsilon$, we assume the form
\begin{align}
\rho(\theta,Q) &=  \epsilon \rho_1(\theta,Q)+
\epsilon^2\rho_2(\theta,Q)+\cdots,
\label{rho-exp} \\ 
\Omega(Q) &= \epsilon\Omega_1(Q)+\epsilon^2\Omega_2(Q)+\cdots.
\label{Omega-exp}
\end{align}
Furthermore, for both $\rho_i(\theta,Q)$ and $\Omega_i(Q)$,
we consider an expansion with regard to $\lambda$. In summary, 
we analyze 
Eq. (\ref{Q-rho}) with  Eqs. (\ref{multi:jint}), (\ref{M-exp}), 
(\ref{rho-exp}), (\ref{Omega-exp}) and $\mu=\epsilon$ and derive 
$\rho_i(\theta,Q)$ and $\Omega_i(Q)$ iteratively. 
These calculations are shown in 
Secs. \ref{no-interacting} and \ref{weak-interacting}.

In order to make our perturbative calculation concrete, we introduce 
a space ${\cal F}$ consisting of all complex-valued, 
square-integrable, periodic functions of $\theta$ on the interval $[0,\ell]$. We endow 
this space with the inner product 
\begin{eqnarray}
(\alpha,\beta)=\int_0^\ell \frac{{\mathrm d}\theta}{\ell}  
\alpha^*(\theta)\beta(\theta)
\end{eqnarray}
for $\alpha, \beta \in {\cal F}$, 
where $^*$ denotes the complex conjugation.
Since the operator $\hat{M}^{(0)}$ is a linear map from ${\cal F}$ 
to ${\cal F}$, we can define all the eigenvalues $\lambda_j$ and 
the corresponding eigenfunctions $\Phi_j (\theta)$ of the operator 
$\hat{M}^{(0)}$ in $\mathcal{F}$ by the equation
\begin{equation}
\hat{M}^{(0)} \Phi_j(\theta) = \lambda_j \Phi_j(\theta),
\label{Phidef}
\end{equation}
where the index $j = 0, \pm 1, \pm 2, \cdots$ is determined by the 
relations $\lambda_j = \lambda_{-j}^*$ and 
$\lambda_0 = 0 > \mathrm{Re}(\lambda_{\pm 1}) 
> \mathrm{Re}(\lambda_{\pm 2}) > \cdots$.
When a complex eigenvalue is degenerate, the corresponding 
labeling is modified appropriately. 
Because $\hat M^{(0)}$ is not a Hermitian operator, it is convenient to 
introduce the adjoint operator of $\hat M^{(0)}{}^{\dagger}$ through 
the relation
\begin{equation}
\left(\hat M^{(0)}{}^{\dagger} \alpha, \beta\right) 
\equiv \left(\alpha, \hat M^{(0)} \beta\right).
\end{equation}
In this space ${\cal F}$, $\hat{M}^{(0)\dagger}$ is explicitly 
represented as
\begin{eqnarray}
\hat{M}^{(0)\dagger}=\pder{ \ }{\theta}\left(
\frac{\Js}{\ps(\theta)}-\frac{T}{\gamma}\der{\log\ps(\theta)}{\theta}
+\frac{T}{\gamma}\pder{ \ }{\theta}\right).
\label{defofMdagger}
\end{eqnarray}
Note that the set of eigenvalues of $\hat M^{(0)}{}^{\dagger}$ 
is identical to that of $\hat M^{(0)}$. Then, we denote the 
eigenfunctions of $\hat M^{(0)}{}^{\dagger}$ as  $\Psi_j (\theta)$ 
and choose their labeling such that
\begin{equation}
\hat M^{(0)}{}^{\dagger} \Psi_j(\theta) =\lambda_j^* \Psi_j(\theta).
\end{equation}
We can choose these eigenfunctions such that the following hold:
\begin{eqnarray}
(\Psi_i, \Phi_j)&=&  \delta_{ij}, 
\label{ortho1} \\
\sum_{j=-\infty}^\infty 
\frac{\Psi_j^*(\theta) \Phi_j(\theta')}{\ell}&=&  \delta(\theta-\theta').
\end{eqnarray}
In particular, 
we choose the zero eigenfunctions as
\begin{align}
\Psi_0(\theta) &= \ps(\theta) , 
\label{zero1} \\
\Phi_0(\theta) &= 1.
\end{align}

\subsection{non-interacting system}\label{no-interacting}

We investigate the non-interacting system by setting $\lambda=0$
in Eq. (\ref{Q-rho}) with Eq. (\ref{multi:noise}).
Equation (\ref{Q-rho}) can then be rewritten as
\begin{align}
&\Omega+\frac{\delta \rho}{\delta Q}\cdot \Omega(Q)\nonumber\\
&=\hat{M}\left[Q+\rho(\theta,Q)\right]\nonumber\\
&-\epsilon\frac{1}{\ps(\theta)}\sqrt{\frac{T}{\gamma}}
\left(
\pder{ \ }{\theta}\bv{\e}_1+\epsilon\pder{ \ }{\bv{X}}
\right)\nonumber\\
&\cdot
\left(\sqrt{\ps(\theta)\left[Q+\rho(\theta,Q)\right]}
\bar{\bv{\xi}}(\theta,\bv{X},t)\right),
\label{nobasis}
\end{align}
for which we use the expansions given by Eqs. (\ref{M-exp}), (\ref{rho-exp}), 
and (\ref{Omega-exp}).
Selecting all the terms proportional to $\epsilon$ 
in Eq. (\ref{nobasis}), 
we obtain
\begin{align}
\Omega_1(Q)
=&
\hat{M}^{(0)}\rho_1(\theta,Q)+\hat{M}^{(1)}Q \nonumber\\ 
&-\frac{1}{\ps(\theta)}\sqrt{\frac{T}{\gamma}}\pder{ \ }{\theta}
\left(\sqrt{\ps(\theta)Q(\bv{X},t)}
\bar{\xi}_1(\theta,\bv{X},t)\right).
\label{ep1}
\end{align}
Here, $\Omega_1(Q)$ and $\rho_1(\theta,Q)$ are unknown
functions. In order to derive them, we set 
\begin{align}
s_1&(\theta, Q)\nonumber\\
\equiv&\Omega_1(Q)-\hat{M}^{(1)}Q\nonumber\\
&+\frac{1}{\ps(\theta)}\sqrt{\frac{T}{\gamma}}\pder{ \ }{\theta}
\left(\sqrt{\ps(\theta)Q(\bv{X},t)}
\bar\xi_1(\theta,\bv{X},t)\right),
\label{b1}
\end{align}
and rewrite Eq. (\ref{ep1}) as 
\begin{equation}
\hat{M}^{(0)}\rho_1(\theta,Q)=s_1(\theta,Q).
\label{lin}
\end{equation}
We can regard
Eq. (\ref{lin}) as a linear equation with respect to $\rho_1$
because $s_1(\theta,Q)$ does not contain $\rho_1$.

Since $\hat{M}^{(0)}$ has a zero eigenvalue,
$\hat{M}^{(0)}$ is not invertible. In this case
there is no unique solution of $\rho_1$ to Eq. (\ref{lin}) 
but either no solution or an infinite number of solutions. 
Then, in order to perform the perturbative calculation 
consistently, we impose the solvability condition 
\begin{eqnarray}
(\Psi_0,s_1)=0;
\label{solv}
\end{eqnarray}
under this condition, 
there exist solutions with an arbitrary constant.
This condition determines $\Omega_1(Q)$ as
\begin{align}
\Omega_1&(Q)\nonumber\\
=&\left(\Psi_0,\hat{M}^{(1)}Q\right)\nonumber\\
&-\left(\Psi_0,\frac{1}{\ps}\sqrt{\frac{T}{\gamma}}
\partial\left[\sqrt{\ps Q(\bv{X},t)}
\bar\xi_1(\cdot,\bv{X},t)\right]\right),
\label{omega1}
\end{align}
where $\partial$ represents the partial derivative with respect to 
$\theta$. We note that the second term in the right-hand side 
of Eq. (\ref{omega1}) 
is equal to zero (see Appendix \ref{treatmulti}).
Then, substituting Eq. (\ref{m1}) into Eq. (\ref{omega1}), we obtain
\begin{eqnarray}
\Omega_1(Q)=-\Js \pder{Q}{X_1}.
\label{omega1non}
\end{eqnarray}

Under the solvability condition given by Eq. (\ref{solv}), 
we can derive the following solutions
of the linear equation expressed by Eq. (\ref{lin}):
\begin{align}
\rho_1&(\theta,Q)\nonumber\\
=&
\sum_{n\not = 0}\frac{\Phi_n(\theta)}{-\lambda_n}
\left(\Psi_n,-\frac{\Js}{\ps}+\frac{T}{\gamma}\partial(\log \ps)\right)
\pder{Q}{X_1}\nonumber\\
&+\sum_{n\not = 0}\frac{\Phi_n(\theta)}{\lambda_n}
\sqrt{\frac{TQ}{\gamma}}
\left(
\Psi_n,\frac{1}{\ps}\partial\left[
\sqrt{\ps}\bar\xi_1(\cdot ,\bv{X},t)\right]\right)
\nonumber\\
&+\chi \Phi_0(\theta).
\label{rho1non}
\end{align}
Here, $\chi$ is an arbitrary constant. 
We set $\chi=0$ hereafter.

Next, we will determine $\Omega_2(Q)$ and $\rho_2(\theta,Q)$. 
Using the terms proportional to $\epsilon^2$ in Eq. (\ref{nobasis}),
we obtain 
\begin{align}
\Omega_2&(Q)+\frac{\delta \rho_1}{\delta Q}\cdot \Omega_1(Q)\nonumber\\
=&\hat{M}^{(0)}\rho_2+\hat{M}^{(1)}\rho_1+\hat{M}^{(2)}Q\nonumber\\
&-\frac{1}{\ps(\theta)}\sqrt{\frac{T}{\gamma}}
\pder{ \ }{\theta}
\left(\sqrt{\ps(\theta)}\frac{\rho_1}{2\sqrt{Q}}
\bar\xi_1(\theta,\bv{X},t)\right) \nonumber \\
&-\frac{1}{\ps(\theta)}\sqrt{\frac{T}{\gamma}}
\pder{ \ }{\bv{X}}\cdot
\left(\sqrt{\ps(\theta)Q}\bar{\bv{\xi}}(\theta,\bv{X},t)\right).
\label{ep2}
\end{align}
In the same manner as that in the first order calculation,
we impose the solvability condition for the linear 
equation of $\rho_2$. This yields 
\begin{align}
\Omega_2&(Q)\nonumber\\
=&\left(\Psi_0,\hat{M}^{(1)}\rho_1\right)
+\left(\Psi_0,\hat{M}^{(2)}Q\right)\nonumber\\
&-\sqrt{\frac{T}{\gamma}}\pder{ \ }{\bv X}\cdot
\left[\sqrt{Q}
\left(\Psi_0,\frac{1}{\sqrt{\ps}}\bar{\bv{\xi}}(\cdot,\bv{X},t)\right)
\right].
\label{omega2non}
\end{align}
Using Eqs. (\ref{m1}) and (\ref{m2}), we obtain 
\begin{align}
\Omega_2(Q)=&D\frac{\partial^2 Q}{\partial X_1^2}
+\frac{T}{\gamma}\frac{\partial^2 Q}{\partial X_2^2}\nonumber\\
&-\pder{ \ }{X_1}\sqrt{Q}\left[\zeta(\bv{X},t)+\eta_1(\bv{X},t)
\right]
\nonumber\\
&-\pder{ \ }{X_2}\sqrt{Q}\eta_2(\bv{X},t).
\label{second:result}
\end{align}
Here, we calculate  $D$, $\zeta(\bv{X},t)$, and $\eta_\alpha(\bv{X},t)$ as
\begin{align}
D=&-\left(b,\left[-\frac{\Js}{\ps}+\frac{T}{\gamma}\partial(\log\ps)\right]
\right)+\frac{T}{\gamma},
\label{result:D} \\
\zeta(\bv{X},t)=&
\sqrt{\frac{T}{\gamma}}
\left(\partial\left(\frac{b}{\ps}\right),
\sqrt{\ps}\bar{\xi}_1(\cdot,\bv{X},t)\right),
\label{result:zeta} \\
\eta_{\alpha}(\bv{X},t)=&
\sqrt{\frac{T}{\gamma}}
\left(\Psi_0,\frac{1}{\sqrt{\ps}}\bar{\xi}_\alpha(\cdot, \bv{X},t)\right),
\label{result:eta} 
\end{align}
where $b(\theta)$ is defined as
\begin{align}
&b(\theta)\nonumber\\
&\equiv \sum_{m\neq 0}
\left(\Psi_0,\left[
-\frac{\Js}{\ps}+\frac{T}{\gamma}\partial(\log \ps)
+\frac{2T}{\gamma}\partial\right]\Phi_m^*\right)
\frac{\Psi_m(\theta)}{\lambda_m^*}.
\label{bdef}
\end{align}
We find that $b(\theta)$ is a real function 
(see Eq. (\ref{result:b})).

Now, we write the coarse-grained hydrodynamic equation by defining
\begin{align}
\tilde Q(\bv{x},t) & \equiv  Q(\bv{X},t), 
\label{coarse1}\\
\Xi_1(\bv{x},t) & \equiv  \frac{1}{\epsilon}\left[\zeta(\bv{X},t)+\eta_1(\bv{X},t)\right],
\label{coarse2}\\
\Xi_2(\bv{x},t) & \equiv  \frac{1}{\epsilon}\eta_2(\bv{X},t).
\label{coarse3}
\end{align}
Using these and from Eqs. (\ref{Qev-gen}), (\ref{omega1non}), 
and (\ref{second:result}), we obtain 
\begin{align}
\pder{\tilde Q}{t}=&-\sum_{\alpha=1}^2 
\pder{\tilde J_\alpha(\bv{x},t)}{x_\alpha},
\label{noncont}
\end{align}
with
\begin{align}
\tilde J_1(\bv{x},t)
&=\Js \tilde Q(\bv{x},t)-
D\pder{\tilde Q}{x_1}+\sqrt{\tilde Q(\bv{x},t)}\Xi_1(\bv{x},t),  \nonumber\\
\tilde J_2(\bv{x},t)
&=-\frac{T}{\gamma}\pder{\tilde Q}{x_2}+
\sqrt{\tilde Q(\bv{x},t)}\Xi_2(\bv{x},t),
\label{nonj}
\end{align}
where $\Xi_{\alpha}$ with $\alpha=1,2$ satisfies
\begin{equation}
\bra \Xi_\alpha(\bv{x},t)\Xi_\beta(\bv{x}',t')\ket 
= 2 B_{\alpha \beta} \delta^2(\bv{x}-\bv{x}')\delta(t-t').
\label{devnon}
\end{equation}
The noise intensities $B_{\alpha\beta}$ are calculated as
$B_{12}=B_{21}=0$ and 
\begin{align}
B_{11}=& \frac{T}{\gamma}\int_0^\ell\frac{{\mathrm d}\theta}{\ell}
\ps(\theta)\left[\frac{ d }{d\theta}
\left(\frac{b(\theta)}{\ps(\theta)}\right)+1\right]^2,
\label{r11} \\
B_{22}=& \frac{T}{\gamma}.
\label{B22D}
\end{align}
See Appendix \ref{D=D} for the calculation.

Here, we present two remarks on the coarse-grained hydrodynamic equation
given by Eq. (\ref{noncont}) with Eqs. (\ref{nonj}) and (\ref{devnon}).
The first remark is on the expression of $D$ given by  
Eq. (\ref{result:D}). Although the expression is complicated, 
we can rewrite Eq. (\ref{result:D}) as
\begin{equation}
D= \frac{T}{\gamma}
\left(\int_0^\ell \frac{{\mathrm d}\theta}{\ell} I_-(\theta)\right)^{-3}
\int_0^\ell \frac{{\mathrm d}\theta}{\ell}
\left[I_-(\theta)\right]^2 I_+(\theta)
\label{Df}
\end{equation}
using Eq. (\ref{Ipm:def}). 
The derivation is presented in Appendix \ref{D=D}. This 
expression of the diffusion constant coincides with that of
the diffusion constant of a Brownian particle in the tilted
periodic potential \cite{Reimann1,Reimann2,HS2}.
Physically, this coincidence is obvious, 
because we consider the non-interacting particles in this subsection.

The second remark is on a special relation
\begin{align}
B_{11} = D. 
\label{B11D}
\end{align}
The proof of this relation is presented in Appendix \ref{D=D}.
This relation corresponds to the fluctuation-dissipation relation 
of the second kind in this fluctuating hydrodynamic equation. 
Using this property,
in the same manner as that in Sec. \ref{model},
we can prove the detailed balance condition of the system
in the moving frame
with velocity $\Js$.
From this condition, we find the steady probability distribution
functional of the coarse-grained density field $P_{\rm S}(Q)$ as
\begin{align}
P_{\rm S}(Q)=\frac{1}{Z_{\rm c}}
\exp\left(
-\int {\mathrm d}^2\bv x Q(\bv{x})
\left[\log Q(\bv x)-1\right]
\right).
\end{align}
This implies that in the non-interacting system,
the density field does not exhibit a long-range spatial correlation
even if the system is out of equilibrium.

\subsection{weakly interacting system}\label{weak-interacting}

In this subsection, we extend the analysis in the previous subsection 
to a system consisting of interacting particles. 
Concretely, 
we expand $\rho_i(\theta,Q)$ and $\Omega_i(Q)$ 
in Eqs. (\ref{rho-exp}) and (\ref{Omega-exp}) as
\begin{align}
\rho_{i}(\theta,Q)=&\rho_{i0}(\theta,Q)+\lambda\rho_{i1}(\theta,Q)
+\lambda^2\rho_{i2}(\theta,Q)
+\cdots,
\label{rho-exp2} \\
\Omega_i(Q)=&\Omega_{i0}(Q)+\lambda \Omega_{i1}(Q)
+\lambda^2 \Omega_{i2}(Q)
+\cdots.
\label{Omega-exp2}
\end{align}
Further, in order to make the calculation results explicit,
we consider two asymptotic cases: (i) 
$\ell_{\rm int}\gg \ell$ and (ii) $\ell_{\rm int}\ll \ell$. 
By substituting Eqs. (\ref{rho-exp2}) and (\ref{Omega-exp2}) 
into Eq. (\ref{Q-rho}) 
with Eq. (\ref{jint}) for the two cases, we determine
$\rho_{ik}(\theta,Q)$ and $\Omega_{ik}$ iteratively.

\subsubsection{case (i)}

We study the case $\ell_{\rm int}\gg\ell$. Specifically, 
we assume that $\ell_{\rm int}\simeq O(\epsilon^{-1} \ell)$, 
and then we set 
\begin{align}
\bar u(\theta,\bv{X})=\epsilon^2 u_{\rm L}(\bv{X}).
\label{longu}
\end{align}
The factor $\epsilon^2$ is introduced in order to develop
a systematic perturbation method. 
Substituting Eq. (\ref{longu}) into Eq. (\ref{multi:jint}),
we express the current generated by this interaction potential as
\begin{align}
\bar{\bv{j}}_{\rm int}&(\theta,\bv{X},t)\nonumber\\
=&\frac{1}{\gamma}\ps(\theta)
\left[Q(\bv{X},t)+\rho(\theta,Q)\right]\nonumber\\
&\int {\mathrm d}^2\bv{Y}
\epsilon\pder{u_{\rm L}(\bv{X}-\bv{Y})}{\bv{X}}Q(\bv{Y},t).
\label{longcu}
\end{align}
Then, Eq. (\ref{Q-rho}) with Eq. (\ref{multi:jint}) becomes
\begin{align}
\Omega&(Q)+\frac{\delta \rho}{\delta Q}\cdot \Omega(Q)
=\hat{M}\left[Q+\rho(\theta,Q)\right]\nonumber \\
&-\epsilon \frac{1}{\ps(\theta)}\sqrt{\frac{T}{\gamma}}
\left(
\pder{ \ }{\theta}\bv{\e}_1+\epsilon\pder{ \ }{\bv{X}}
\right)\nonumber\\
&\cdot \left\{\sqrt{\ps(\theta)
\left[Q+\rho(\theta,Q)\right]}\bar{\bv{\xi}}(\theta,\bv{X},t)\right\}
\nonumber \\
&-\frac{\lambda}{\gamma}\epsilon\frac{1}{\ps(\theta)}
\left(
\pder{ \ }{\theta}\bv{\e}_1+\epsilon\pder{ \ }{\bv{X}}
\right)\nonumber\\
&\cdot\left\{
\ps(\theta)\left[Q+\rho(\theta,Q)\right]
\int {\mathrm d}^2\bv{Y}\pder{u_{\rm L}(\bv{X}-\bv{Y})}{\bv{X}}Q(\bv{Y},t)
\right\}.
\label{basislo}
\end{align}
We now  analyze Eq. (\ref{basislo}) with the expansions given by
Eqs. (\ref{M-exp}), (\ref{rho-exp}), (\ref{Omega-exp}), 
(\ref{rho-exp2}), and (\ref{Omega-exp2}).

Selecting all the terms proportional to $\epsilon$, 
we obtain 
\begin{eqnarray}
\Omega_1(Q)=&&\hat{M}^{(0)}\rho_1+\hat{M}^{(1)}Q\nonumber\\
&-&\frac{1}{\ps(\theta)}\sqrt{\frac{T}{\gamma}}
\pder{ \ }{\theta}
\left[\sqrt{\ps(\theta)Q}\bar{\xi}_1(\theta, \bv{X},t)\right]
\nonumber\\
&-&\frac{\lambda}{\gamma}\frac{1}{\ps(\theta)}\pder{\ps(\theta)}{\theta}
Q(\bv{X},t)\nonumber\\
&&\int {\mathrm d}^2\bv{Y}
\pder{u_{\rm L}(\bv{X}-\bv{Y})}{X_1}Q(\bv{Y},t).
\label{long1}
\end{eqnarray}
The terms independent of $ \lambda$ reproduce Eq. (\ref{ep1}).
Thus, $\Omega_{10}(Q)$ and $\rho_{10}$ are equal to $\Omega_1(Q)$
and $\rho_1$ given by Eqs. (\ref{omega1non}) and (\ref{rho1non}), 
respectively. 
Next, extracting all the terms proportional to $\lambda$ 
in Eq. (\ref{long1}), 
we obtain 
\begin{eqnarray}
\hat{M}^{(0)}\rho_{11}=&&\Omega_{11}(Q)\nonumber\\
&+&\frac{1}{\gamma}\frac{1}{\ps(\theta)}
\pder{\ps(\theta)}{\theta}
Q(\bv{X},t)\nonumber\\
&&\int {\mathrm d}^2\bv{Y}\pder{u_{\rm L}(\bv{X}-\bv{Y})}{X_1}Q(\bv{Y},t).
\label{long11}
\end{eqnarray}
The solvability condition for the linear equation of $\rho_{11}$ yields
\begin{align}
\Omega_{11}(Q)=0.
\label{loomega11}
\end{align}
Then, we can solve $\rho_{11}$ as
\begin{align}
\rho_{11}(\theta,Q)=&
\frac{1}{\gamma}
\sum_{n\neq 0}
\frac{\Phi_n(\theta)}{\lambda_n}
\left(\Psi_n,\frac{1}{\ps}\partial \ps\right)
Q(\bv{X},t)\nonumber\\
&\int {\mathrm d}^2\bv{Y}\pder{u_{\rm L}(\bv{X}-\bv{Y})}{X_1}
Q(\bv{Y},t),
\label{lorho11}
\end{align}
where the term proportional to $\Phi_0(\theta)$ is set to zero.

Next, using all the terms proportional to $\epsilon^2$ 
in Eq. (\ref{basislo}) with the expansions,
we calculate $\Omega_{20}(Q)$ and $\Omega_{21}(Q)$ by repeating the 
same analysis. Obviously, $\Omega_{20}(Q)$ is equal to $\Omega_2(Q)$ 
in Eq. (\ref{second:result}).
Extracting the term proportional to $\lambda\epsilon^2$ in Eq. (\ref{basislo}),
we obtain
\begin{align}
\Omega_{21}&(Q)
+\frac{\delta \rho_{11}}{\delta Q}\cdot \Omega_{10}(Q)
+\frac{\delta \rho_{10}}{\delta Q}\cdot \Omega_{11}(Q)\nonumber\\
=&\hat{M}^{(0)}\rho_{21}+\hat{M}^{(1)}\rho_{11}\nonumber\\
&-\frac{1}{\ps(\theta)}\sqrt{\frac{T}{\gamma}}
\pder{ \ }{\theta}
\left[\sqrt{\ps(\theta)}
\frac{\rho_{11}(\theta,Q)}{2\sqrt{Q(\bv{X},t)}}
\bar\xi_1(\theta,\bv{X},t)\right]\nonumber\\
&-\frac{1}{\gamma}\frac{1}{\ps(\theta)}
\pder{ \ }{\theta}\left[\ps(\theta) \rho_{10}(\theta,Q)
\right]\nonumber\\
&\int {\mathrm d}^2\bv{Y}\pder{u_{\rm L}(\bv{X}-\bv{Y})}{X_1}
Q(\bv{Y},t)\nonumber\\
&-\frac{1}{\gamma}\pder{ \ }{\bv{X}}\cdot\left[
Q(\bv{X},t)\int {\mathrm d}^2\bv{Y}
\pder{u_{\rm L}(\bv{X}-\bv{Y})}{\bv{X}}Q(\bv{Y},t)\right].
\label{long21}
\end{align}
Then, the solvability condition for the linear equation of $\rho_{21}$ yields
\begin{align}
\Omega&_{21}(Q)\nonumber\\
=&\left(\Psi_0,\hat{M}^{(1)}\rho_{11}\right)\nonumber\\
&-\frac{1}{\gamma}\pder{ \ }{\bv{X}}\cdot
\left(
Q(\bv{X},t)\int {\mathrm d}^2\bv{Y}\pder{u_{\rm L}(\bv{X}-\bv{Y})}{\bv{X}}
Q(\bv{Y},t)
\right).
\label{ap:loomega21}
\end{align}
Substituting Eq. (\ref{lorho11}) into 
the first term in the right-hand side of Eq. (\ref{ap:loomega21}) leads to
\begin{align}
\Big(\Psi_0&\left.,\hat{M}^{(1)}\rho_{11}\right)\nonumber\\
=&\frac{1}{\gamma}\left(b,\frac{1}{\ps}\partial\ps\right)\nonumber\\
&\pder{ \ }{X_1}\left(Q(\bv{X},t)\int {\mathrm d}^2\bv{Y}
\pder{u_{\rm L}(\bv{X}-\bv{Y})}{X_1}Q(\bv{Y},t)\right),
\label{ap:lom1rho11}
\end{align}
where $b(\theta)$ is given by Eq. (\ref{bdef}).
Here, we note the identity
\begin{align}
\Big(b,&\frac{1}{\ps}\partial\ps\Big)-1\nonumber\\
&=-\int_0^\ell\frac{{\mathrm d}\theta}{\ell}\ps(\theta)\left(
\pder{ \ }{\theta}\left[\frac{b(\theta)}{\ps(\theta)}\right]+1\right)
\nonumber\\
&=-\left(\int_0^\ell\frac{{\mathrm d}\theta'}{\ell}I_-(\theta')\right)^{-2}
\int_0^\ell\frac{{\mathrm d}\theta}{\ell}I_-(\theta)I_+(\theta)
\nonumber\\
&=-\gamma\der{\Js}{f},
\label{diffmovi}
\end{align}
where we have used Eq. (\ref{partialb3}) to obtain the third line,
and the fourth line can be confirmed directly from Eq. (\ref{defofJs})
(see also Ref. \cite{HS2}).
Using Eqs. (\ref{ap:lom1rho11}) and (\ref{diffmovi}),
we rewrite Eq. (\ref{ap:loomega21}) as
\begin{align}
\Omega&{}_{21}(Q)\nonumber\\
=&-\der{\Js}{f}
\pder{ \ }{X_1}
\left[
Q(\bv{X},t)\int {\mathrm d}^2\bv{Y}\pder{u_{\rm L}(\bv{X}-\bv{Y})}{X_1}
Q(\bv{Y},t)\right]
\nonumber\\
&-\frac{1}{\gamma}\pder{ \ }{X_2}
\left[
Q(\bv{X},t)\int {\mathrm d}^2\bv{Y}\pder{u_{\rm L}(\bv{X}-\bv{Y})}{X_2}
Q(\bv{Y},t)\right].
\label{loomega21}
\end{align}

Finally, we derive the coarse-grained hydrodynamic  equation 
from Eqs. (\ref{Qev-gen}), (\ref{omega1non}), (\ref{second:result}), 
(\ref{loomega11}), and (\ref{loomega21}). 
Using the variable defined in 
Eqs. (\ref{coarse1}), (\ref{coarse2}), and (\ref{coarse3}), 
we obtain the continuity equation for $\tilde Q(\bv{x},t)$ 
expressed by Eq. (\ref{noncont}) 
with the current as follows:
\begin{align}
\tilde J_1&(\bv{x},t)
=\Js \tilde Q(\bv{x},t)
-D\pder{\tilde{Q}}{x_1}+\sqrt{\tilde Q(\bv{x},t)}\Xi_1(\bv{x},t)\nonumber\\
&+\lambda\frac{D}{T}(1-\delta)\tilde{Q}(\bv{x},t)
\int {\mathrm d}^2\bv{y}\pder{u(\bv{x}-\bv{y})}{x_1}\tilde{Q}(\bv{y},t),
\nonumber\\
\tilde J_2&(\bv{x},t)
=-\frac{T}{\gamma}\pder{\tilde{Q}}{x_2}
+\sqrt{\tilde Q(\bv{x},t)}\Xi_2(\bv{x},t)\nonumber\\
&+\lambda\frac{1}{\gamma}\tilde{Q}(\bv{x},t)
\int {\mathrm d}^2\bv{y}\pder{u(\bv{x}-\bv{y})}{x_2}\tilde{Q}(\bv{y},t),
\label{long:result}
\end{align}
where $\delta$ is the dimensionless parameter defined as
\begin{align}
\delta\equiv 1-\frac{T}{D}\der{\Js}{f}.
\label{defofdelta}
\end{align}
The Einstein relation leads to $\delta=0$
when $f=0$, 
while $\delta\neq 0$ when $f\neq 0$ as far as we 
checked numerically (see Ref. \cite{HS2}).

When $\delta=0$, 
by using a similar argument as that in Sec. \ref{model},
we can prove the detailed balance condition
of the system in the moving frame with velocity $\Js$.
However, when $\delta\neq 0$,
the coarse-grained hydrodynamic equation does not possess
the detailed balance property for any moving frame
because we cannot construct a potential function
such as $H$ for the argument in Sec. \ref{model}.
It should be noted that the noise intensities 
are not modified by the lowest order contribution of the interaction effects. 
Therefore, the fluctuation-dissipation 
relation of the second kind is maintained 
in this hydrodynamic equation.

\subsubsection{case (ii)}\label{shortrange}

Next, we study the case $\ell_{\rm int}\ll\ell$. 
We assume the form 
\begin{align}
\bar u(\theta,\bv{X})
=u_0\ell\delta(\theta)\epsilon^2\delta^2(\bv{X}),
\label{shortu}
\end{align}
where the intensity $u_0$ is determined  from the interaction 
potential with  $\ell_{\rm int}\ll \ell$.
Here, we set $u_0= \epsilon u_{\rm S}$
in order to develop a systematic perturbation.
Note that the quantity $u_0$ should appear
when we calculate experimentally measurable quantities.
Substituting Eq. (\ref{shortu}) into Eq. (\ref{multi:jint}), 
we obtain 
\begin{align}
\bar{\bv{j}}_{\rm int}&(\theta,\bv{X},t)\nonumber\\
=&\epsilon \frac{u_{\rm S}}{\gamma}
\ps(\theta)\left[Q+\rho(\theta,Q)\right]\nonumber\\
&\left(
\pder{ \ }{\theta}\bv{\e}_1+\epsilon\pder{ \ }{\bv{X}}
\right)\left\{
\ps(\theta)\left[Q+\rho(\theta,Q)\right]\right\}.
\label{shortcu}
\end{align}

Then, Eq. (\ref{Q-rho}) with Eq. (\ref{multi:jint}) becomes
\begin{align}
\Omega(&Q)+\frac{\delta \rho}{\delta Q}\cdot \Omega(Q)
=\hat{M}\left[Q+\rho(\theta,Q)\right]\nonumber\\
-&\mu\sqrt{\frac{T}{\gamma}}\frac{1}{\ps(\theta)}
\left(
\pder{ \ }{\theta}\bv{\e}_1+\epsilon\pder{ \ }{\bv{X}}
\right)\nonumber\\
&\cdot\left[\sqrt{\ps(\theta)\left[Q+\rho(\theta,Q)\right]}
\bar{\bv{\xi}}(\theta,\bv{X},t)\right]
\nonumber\\
-&\epsilon\frac{\lambda u_{\rm S}}{\gamma}
\frac{1}{\ps(\theta)}
\left(
\pder{ \ }{\theta}\bv{\e}_1+\epsilon\pder{ \ }{\bv{X}}
\right)\nonumber\\
\cdot&\Big\{\ps(\theta)\left[Q+\rho(\theta,Q)\right]\nonumber\\
&\left(
\pder{ \ }{\theta}\bv{\e}_1+\epsilon\pder{ \ }{\bv{X}}
\right)
\ps(\theta)\left[Q+\rho(\theta,Q)\right]\Big\}.
\label{basissh}
\end{align}
We analyze Eq. (\ref{basissh}) using the expansions given 
in Eqs. (\ref{M-exp}), (\ref{rho-exp}), (\ref{Omega-exp}), (\ref{rho-exp2}),
and (\ref{Omega-exp2}).

The calculation procedures hereafter
are the same as that for case (i).
Thus, without repeating the calculations, 
we summarize the results as follows:
\begin{align}
\Omega_{11}(Q)&=0,\label{shomega11}\\
\rho_{11}(\theta,Q)&=
Q^2\frac{u_{\rm S}}{\gamma}
\sum_{n\neq 0}\frac{\Phi_n(\theta)}{\lambda_n}\nonumber\\
&\int_0^\ell \frac{{\mathrm d}\theta'}{\ell}
\frac{\Psi^*_n(\theta')}{\ps(\theta')}\der{ \ }{\theta'}\left[
\ps(\theta')
\der{ \ps(\theta') }{\theta'}\right],
\label{shrho11}\\
\Omega_{21}(Q)=&-\pder{ \ }{X_1}[\bar\nu Q^2(\bv{X},t)],
\label{shomega21}
\end{align}
where $\bar\nu$ is defined as
\begin{align}
\bar\nu &=\frac{u_{\rm S}}{\gamma}
\left(\int_0^\ell\frac{{\mathrm d}\theta}{\ell}I_-(\theta)\right)^{-1}
\int_0^\ell\frac{{\mathrm d}\theta}{\ell}
\ps(\theta)I_+(\theta)\der{\ps(\theta)}{\theta}.
\end{align}
Note that in the equilibrium case ($f=0$), $\bar \nu=0$
because $\ps(\theta)I_+(\theta)$ is equal to unity when $f=0$.

Finally, we derive the coarse-grained hydrodynamic equation 
from Eqs. (\ref{omega1non}), (\ref{second:result}), (\ref{shomega11}),
and (\ref{shomega21}). Using the variable defined by 
Eqs. (\ref{coarse1}), (\ref{coarse2}), and (\ref{coarse3}),
we obtain  the continuity equation for $\tilde Q(\bv{x},t)$ 
expressed by Eq. (\ref{noncont}) with the current as follows: 
\begin{align}
\tilde J_1(\bv{x},t)
&=\Js\tilde Q(\bv{x},t)
-D\pder{\tilde Q}{x_1}+\sqrt{\tilde Q(\bv{x},t)}\Xi_1(\bv{x},t)\nonumber\\
&+\lambda \nu \tilde Q^2(\bv{x},t), \nonumber \\
\tilde J_2(\bv{x},t)
&=-\frac{T}{\gamma}\pder{\tilde Q(\bv{x},t)}{x_2}+\sqrt{\tilde Q(\bv{x},t)}
\Xi_2(\bv{x},t).
\label{short:result}
\end{align}
Here, we have defined $\nu$ as 
\begin{align}
\nu\equiv&\frac{u_0}{u_{\rm S}}\bar \nu,
\label{defofc}
\end{align}
where 
the quantity $\nu$ is independent of $\epsilon$
and its value can be determined from the Langevin model we study.
Note that the coarse-grained hydrodynamic equation 
does not possess the detailed balance property for any moving frame
except for the case where $\nu=0$.

\subsection{correlation function}\label{correlation}

In this subsection, we calculate the equal-time correlation function 
$C_0(\bv{r})$
of the coarse-grained density field for the two above-mentioned cases, 
case (i) $\ell_{\rm int}\gg \ell$ and case (ii) $\ell_{\rm int}\ll \ell$.
We first define the space-time correlation 
function as 
\begin{align}
C(\bv{r},\tau)\equiv\bra \psi(\bv{x},t)\psi(\bv{x}+\bv{r},t+\tau)\ket,
\label{cdef}
\end{align}
where $\psi(\bv{x},t)$ represents the deviation of the density field 
from the average value $\bar \rho$ as follows:
\begin{align}
\psi(\bv{x},t)\equiv \tilde Q(\bv{x},t)-\bar\rho.
\label{defofpsi}
\end{align}
In Eq. (\ref{cdef}), we assume that the
statistical properties of $\psi(\bv{x},t)$ are 
translational invariant in the space and time directions.

Hereinafter, we denote the Fourier-transformation of a 
function $f(\bv{x},t)$ as
\begin{align}
\hat{f}(\bv{k},\omega)=&
\int{\mathrm d}^2\bv{x}{\mathrm d}t
f(\bv{x},t)\e^{-i\bv{k}\cdot\bv{x}-i\omega t}.
\label{Fourier}
\end{align}
We also use the same notation for the Fourier-transformation of a 
function $f(\bv{x})$ as follows:
\begin{align}
\hat f(\bv{k})=\int {\mathrm d}^2\bv{x}f(\bv{x})\e^{-i\bv{k}\cdot\bv{x}}.
\end{align}
Then, we can derive the relation 
\begin{align}
\hat{C}(\bv{k},\omega)(2\pi)^3\delta^2(\bv{k}+\bv{k}')\delta(\omega+\omega')
=\bra\hat{\psi}(\bv{k},\omega)\hat{\psi}(\bv{k'},\omega')\ket.
\end{align}
Further, 
the equal-time correlation function $C_0(\bv{r})$ is given by
\begin{align}
C_0(\bv{r}) 
=& \int \frac{{\mathrm d}^2 \bv{k}}{(2\pi)^2} 
\hat C_0(\bv{k})\e^{i\bv{k}\cdot\bv{r}}\nonumber\\
=& \int \frac{{\mathrm d}^2 \bv{k}}{(2\pi)^2} 
\frac{{\mathrm d} \omega}{2\pi}\hat C(\bv{k},\omega)\e^{i\bv{k}\cdot\bv{r}}.
\label{c0def}
\end{align}
In the argument below, we first calculate $\hat C(\bv{k},\omega)$ 
from the coarse-grained hydrodynamic equations for cases (i) and (ii),
and we derive $C_0(\bv{r})$ using Eq. (\ref{c0def}).

\subsubsection{case (i)}

Let us consider the case $\ell_{\rm int} \gg \ell$. 
Substituting Eq. (\ref{defofpsi}) into the continuity equation 
expressed by Eq. (\ref{noncont}) 
with Eq. (\ref{long:result}),
we can obtain the equation for $\psi$. We further linearize 
the obtained equation. Then, the resultant equation becomes
\begin{align}
\pder{\psi}{t}=-\sum_{\alpha=1}^2
\pder{{\cal J}_\alpha}{x_\alpha},
\label{contpsi}
\end{align}
with the current ${\cal J}_\alpha$ expressed as
\begin{align}
{\cal J}_1(\bv{x},t)
=&\Js \psi-D\pder{ \psi}{x_1}\nonumber\\
&-D\frac{\lambda\bar\rho(1-\delta)}{T}
\int {\mathrm d}^2\bv{y}\pder{u(\bv{x}-\bv{y})}{x_1} \psi(\bv{y},t)
\nonumber\\
&+\sqrt{\bar\rho}\Xi_1(\bv{x},t),\nonumber\\
 {\cal J}_2(\bv{x},t)
=&-\frac{T}{\gamma}\pder{ \psi}{x_2}
-\frac{\lambda\bar\rho}{\gamma}
\int {\mathrm d}^2\bv{y}\pder{u(\bv{x}-\bv{y})}{x_2} \psi(\bv{y},t)
\nonumber\\
&+\sqrt{\bar\rho}\Xi_2(\bv{x},t).
\end{align}

The Fourier-transform of the evolution equation is written as
\begin{align}
\hat{\psi}(\bv{k},\omega)=G(\bv{k},\omega)
\left(-i\sqrt{\bar\rho}\bv{k}\cdot\hat{\bv{\Xi}}(\bv{k},\omega)
\right).
\end{align}
Here, $G(\bv{k},\omega)$ is the Green function calculated as
\begin{align}
\frac{1}{G(\bv{k},\omega)}
=&i(\omega+\Js k_1)
+g_\delta(\bv{k})Dk_1^2+g_0(\bv{k})\frac{T}{\gamma}k_2^2,
\label{GreenFunctionL}
\end{align}
where we have defined $g_\delta(\bv{k})$ as 
\begin{align}
g_\delta(\bv k)\equiv
1+\frac{\lambda\bar\rho\hat u(\bv k)}{T}(1-\delta).
\end{align}

Using the relation
\begin{align}
\bra\hat\Xi_\alpha(\bv{k},\omega)\hat\Xi_\beta(\bv{k}',\omega')\ket
=(2\pi)^32B_{\alpha\beta}\delta^2(\bv{k}+\bv{k}')\delta(\omega+\omega'),
\label{defofnoise:omegak}
\end{align}
we obtain 
\begin{align}
\hat C(\bv{k},\omega)
=2\bar\rho \left|G(\bv{k},\omega)
\right|^2
\left(Dk_1^2+\frac{T}{\gamma}k_2^2\right).
\label{ckomegalong}
\end{align}
Integrating Eq. (\ref{ckomegalong}) 
with Eq. (\ref{GreenFunctionL}) over the frequency, we calculate 
\begin{align}
\hat C_0(\bv{k})=&
\bar\rho
\frac{g_{\delta/2}(\bv{k})}{g_\delta(\bv{k})g_0(\bv{k})}\nonumber\\
&+
\lambda\bar\rho\delta\frac{\bar\rho\hat u(\bv{k})}{2T}
\frac{1}{g_\delta(\bv{k})g_0(\bv{k})}\nonumber\\
&\times\frac{g_\delta(\bv{k})Dk_1^2-g_0(\bv{k})(T/\gamma)k_2^2}
{g_\delta(\bv{k})Dk_1^2+g_0(\bv{k})(T/\gamma)k_2^2}.
\label{cklonglin}
\end{align}

The asymptotic behavior in the range $|\bv{k}|\ll \ell_{\rm int}^{-1}$ 
in Eq. (\ref{cklonglin}) is evaluated as 
\begin{align}
\hat C_0(\bv{k})\simeq&
\bar\rho
\frac{g_{\delta/2}(0)}{g_\delta(0)g_0(0)}\nonumber\\
&+
\lambda\bar\rho\delta\frac{\bar\rho\hat u(0)}{2T}
\frac{1}{g_\delta(0)g_0(0)}\nonumber\\
&\times\frac{g_\delta(0)Dk_1^2-g_0(0)(T/\gamma)k_2^2}
{g_\delta(0)Dk_1^2+g_0(0)(T/\gamma)k_2^2},
\label{cklongasy}
\end{align}
From this expression, the  asymptotic form of $C_0(\bv{r})$ in the 
range $|\bv{r}|\gg\ell_{\rm int}$ is derived as
\begin{align}
C_0(\bv{r})\simeq&
-\frac{\lambda\bar\rho^2\hat u(0)\delta}
{2\pi T\sqrt{\left[g_\delta(0) g_0(0)\right]^3 DT/\gamma}}
\nonumber\\
&\times\frac
{\left[g_\delta(0)D\right]^{-1}r_1^2-[g_0(0)(T/\gamma)]^{-1}r_2^2}
{\left\{
\left[g_\delta(0)D\right]^{-1}r_1^2+[g_0(0)(T/\gamma)]^{-1}r_2^2
\right\}^2},
\label{lrclong}
\end{align}
where $\bv{r}=(r_1,r_2)$.
This represents a long-range correlation of the type $1/r^2$.

\subsubsection{case (ii)}

Next, let us consider the case $\ell_{\rm int} \ll \ell$. 
In this case, we obtain the continuity equation for $\psi$
with the current ${\cal J}_\alpha$ expressed as
\begin{align}
{\cal J}_1(\bv{x},t)=&
(\Js+2\bar \rho\lambda \nu)\psi-D\pder{\psi}{x_1}
+\sqrt{\bar\rho+\psi(\bv x,t)}\Xi_1(\bv{x},t)
\nonumber\\
&+\lambda \nu\psi^2(\bv{x},t)\nonumber \\
{\cal J}_2(\bv{x},t)=&
-\frac{T}{\gamma}\pder{\psi}{x_2}
+\sqrt{\bar\rho+\psi(\bv x,t)}\Xi_2(\bv{x},t)
\label{short:equivalent}.
\end{align}
Further, for simplicity, we assume that the $\psi$ dependence 
of the noise term can be neglected. 
Then, the  evolution equation coincides with 
the special case of that investigated in Ref. \cite{NOS}, where 
the fluctuation-dissipation relation of the
second kind is satisfied.
According to the result of Ref. \cite{NOS},
the equal-time correlation function is not modified by the interaction effects
up to the second order of $\lambda$ in this case (see Eq. (B2) in Ref. \cite{NOS}).
Thus, we conclude that there is no long-range correlation within this 
approximation.

\section{concluding remark}\label{final}

The main achievement of this study is the derivation of the
coarse-grained fluctuating hydrodynamic equation for the driven many-body 
Langevin system.
In the two asymptotic cases 
for the interaction range between particles, 
which are given by Eqs. (\ref{longu}) and (\ref{shortu}),
we derived the two expressions of particle current, Eqs. (\ref{long:result}) 
and (\ref{short:result}), respectively, with the continuity equation 
expressed by Eq. (\ref{noncont}). 
Using the obtained evolution equations,
we calculated the equal-time correlation function 
of the coarse-grained density field for each case. We found that this
correlation function exhibits 
the long-range correlation of the type $r^{-d}$ 
in the case given by Eq. (\ref{longu}), while 
no such behavior was observed in the case given by Eq. (\ref{shortu}).

We derived the coarse-grained fluctuating hydrodynamic equation by
applying  a singular perturbation method
to a stochastic partial differential equation. 
The method is standard  except for the treatment of space-time noise 
(see Appendix \ref{treatmulti}). 
We expect that our method can
be used to investigate other related problems such as 
phase diffusion behavior in periodic pattern formations 
under the influence of noise. 
We also note that 
the derived 
coarse-grained fluctuating hydrodynamic equations in this study 
contain non-linear functions of $f$ 
such as $\Js$, $D$, $B_\alpha$, $\delta$, and $\nu$.
Therefore, we can discuss the system behavior in a non-linear range 
with respect to $f$.

The long-range correlation we obtained for the case 
given by Eq. (\ref{longu})
has the essentially same mechanism as that of Ref.
\cite{Sasa2particles} which studied the system consisting of 
two Brownian particles under an external force. 
Thus, the result in this paper is regarded as an extension 
of that in Ref. \cite{Sasa2particles},
although the Fokker-Plank equation was analyzed in this reference.
On the other hand, the long distance
behavior for the case given by Eq. (\ref{shortu}) might be strange, because 
it has been speculated  that an anisotropic system without 
the detailed balance condition generically exhibits 
the long-range correlation
\cite{Dorfman}. Here, it is noteworthy that the long-range 
correlation of the type $r^{-d}$ 
does not appear in driven lattice gases 
with the evolution rule called an exponential method, while 
it appears in the cases of a heat bath method and a 
Metropolis method \cite{Tasaki}.
It might be interesting
to find a connection of our result with that reported in 
Ref. \cite{Tasaki}.

Our calculation result for the correlation function 
given by Eq. (\ref{cklonglin}) 
provides the functional form  of its short-range part.
In contrast to the long-range part, the short-range part 
depends on the details of the system such as the selection
of the interaction potential. Such a non-universal part 
has never been investigated intensively. 
Here, let us recall that 
the statistical properties of density fluctuations are described
by the free energy function of the system if the system is
in equilibrium. Therefore, it might be expected that the 
short-range part of the correlation function is related to
a thermodynamic function extended to non-equilibrium steady
states. 
Although thermodynamics in non-equilibrium steady states has 
not yet been established, there exists one promising approach to
construct a consistent framework whose validity can be
checked experimentally \cite{SST}. According to this framework, 
the statistical properties of density fluctuations can be described
by an extended free energy determined operationally when the 
effect of the long-range correlation is removed. Indeed, 
by  numerical experiments on a driven lattice gas, 
it was demonstrated that the intensity 
of density fluctuations of a particular type
is determined by an extended free energy 
\cite{HS1}, and this was proved in Ref. \cite{SST}.  We expect
that a similar analysis can be performed for the Langevin system
under investigation in this study. 
The calculation of the short-range part of the correlation 
function is indispensable in this analysis.

The most ambitious goal is to provide a unified description of density
fluctuations in an elegant manner. Even if the short-range 
part of the statistical property of density fluctuations 
is determined by an extended thermodynamic
function, the long-range correlation that destroys the 
extensive nature of the system is obviously out of 
thermodynamic consideration. Here, it should be noted 
that a variational principle referred to as 
{\it the additivity principle}
is effective to describe the long-range behavior of 
density fluctuations in non-equilibrium lattice gases \cite{DLS1,DLS2}.
Although we do not know a class of models 
to which this principle can be applied,
it is interesting to determine whether this principle 
can be applied to 
the Langevin system under investigation.

The authors acknowledge H. Tasaki and K. Hayashi 
for useful discussions and comments.
This work was supported by a grant from the Ministry of Education,
Science, Sports and Culture of Japan (No. 16540337). 

\appendix
\section{fluctuating hydrodynamics 
for fine-grained density field}\label{deridens}

We derive the evolution equation for a fine-grained density 
field 
from the Langevin equation given by Eq. (\ref{lange}).
Mathematically, the essence of the derivation
is in employing the Ito formula for arbitrary functions
of the fluctuating variable \cite{Gard}, and the evolution equation 
for the fine-grained density field is obtained by a standard
treatment of Dirac's  delta function. This approach was performed 
in Ref. \cite{Dean}.  
Although Ref. \cite{Dean}
provides sufficient information on the derivation
of the evolution equation for a fine-grained density field,
we present a different method for the derivation in this Appendix. 
We find that this  method is less mathematical, but more pedagogical 
than the standard one.

Let $\Delta t$ be a sufficiently small time interval, and set 
$t_n=n\Delta t$. Then, from Eq. (\ref{lange}),
the movement of each particle during the time interval, 
$\Delta x_{i\alpha}(t_n)\equiv x_{i\alpha}(t_{n+1}) -x_{i\alpha}(t_n)$, 
can be expressed as 
\begin{align}
\gamma&\Delta x_{i \alpha}(t_n)
\nonumber\\
=&
\left[
\left(f-\pder{U(x_{i1})}{x_{i1}}\right)
\delta_{1\alpha}
-\sum_{j\neq i}\pder{u(\bv{x}_i -\bv{x}_j)}{x_{i \alpha}}
\right]_{[\bv{x}_\ell=\bv{x}_\ell(t_n)]}\Delta t
\nonumber\\
&+\hat{W}_{i \alpha}(t_n)+O\left((\Delta t)^{3/2}\right),
\label{disx}
\end{align}
where 
\begin{equation}
\hat{W}_{i \alpha}(t_n)=\int_{t_n}^{t_{n+1}} {\mathrm d}t R_{i\alpha}(t),
\end{equation}
and it should be noted that the equality 
\begin{align}
\hat W_{i\alpha}(t_n)\hat W_{j\beta}(t_m)=
2\gamma T \delta_{ij}\delta_{\alpha\beta}\delta_{nm}\Delta t 
\label{ito}
\end{align}
holds almost surely \cite{Gard}.  

Next, for the fine-grained density field $\rhod(\bv{x},t)$  
defined by Eq. (\ref{dens}),  we obtain 
\begin{align}
\gamma&\left[\rhod(\bv{x},t_{n+1})-\rhod(\bv{x},t_n)\right]
\nonumber\\
=&
\sum_{i \alpha}\Delta x_{i \alpha}(t_n)
\pder{ \ }{x_{i \alpha}}\rhod(\bv{x},t_n)\nonumber\\
&+\frac{1}{2}\sum_{i \alpha}\sum_{j \beta}
\Delta x_{i \alpha}(t_n)\Delta x_{j \beta}(t_n)
\pder{ \ }{x_{i \alpha}}\pder{ \ }{x_{j \beta}}\rhod(\bv{x},t_n)\nonumber\\
&+O\left((\Delta t)^{3/2}\right).
\end{align}
Substituting Eq. (\ref{disx}) into the above expression 
and using Eq. (\ref{ito}), 
we derive 
\begin{align}
\gamma&\left[\rhod(\bv{x},t_{n+1})-\rhod(\bv{x},t_n)\right]
\nonumber\\
&=-\Delta t\pder{ \ }{x_1}\left[
\left(f-\pder{U(x_1)}{x_1}\right)\rhod(\bv{x},t)\right]\nonumber\\
&+\Delta t
\pder{ \ }{\bv{x}}\cdot\int {\mathrm d}^2\bv{y}
\rhod(\bv{x},t)\pder{u(\bv{x} -\bv{y})}{\bv{x}}\rhod(\bv{y},t)\nonumber\\
&+T\Delta t\left(\pdert{ \ }{x_1}
+\pdert{ \ }{x_2}\right)\rhod(\bv{x},t)\nonumber\\
&-\pder{ \ }{\bv{x}}\cdot
\bv{W}(\bv{x},t_n)
+O\left((\Delta t)^{3/2}\right),
\label{derivativerhod}
\end{align}
where we have defined 
\begin{align}
W_\alpha(\bv{x},t_n)
\equiv\sum_{i=1}^{N}\hat W_{i \alpha}(t_n)\delta^2(\bv{x}-\bv{x}_i).
\end{align}
Note that $W_\alpha(\bv{x},t_n)$ satisfies
\begin{align}
&\bra W_\alpha(\bv{x},t_n)W_\beta(\bv{x}',t_m)\ket\nonumber\\
&=2\gamma T 
\rhod(\bv{x},t_n)\delta_{\alpha \beta}\delta^2(\bv{x}-\bv{x}')\delta_{mn}\Delta t.
\end{align}

Finally, taking  the limit $\Delta t\rightarrow 0$, we obtain
\begin{align}
\pder{\rhod(\bv{x},t)}{t}
&=-\frac{1}{\gamma}\pder{ \ }{x_1}\left[
\left(-\pder{U(x_1)}{x_1}+f\right)\rhod(\bv{x},t)\right]\nonumber\\
&+\frac{1}{\gamma}
\pder{ \ }{\bv{x}}\cdot\int {\mathrm d}^2\bv{y}
\rhod(\bv{x},t)\pder{u(\bv{x} -\bv{y})}{\bv{x}}\rhod(\bv{y},t)\nonumber\\
&+\frac{T}{\gamma}\left(\pdert{ \ }{x_1}
+\pdert{ \ }{x_2}\right)\rhod(\bv{x},t)
\nonumber\\
&-\pder{ \ }{\bv{x}}\cdot\sqrt{\frac{T\rhod(\bv{x},t)}{\gamma}}
\bv{\xi}(\bv{x},t),
\end{align}
where $\xi_\alpha(\bv{x},t)$ satisfies Eq. (\ref{noise}).
It is easily confirmed that the final expression  is 
equivalent to Eq. (\ref{conti}) with Eq. (\ref{devdens}).

\section{function of $(\theta,\bv{X})$}\label{treatmulti}

In Sec. \ref{basicframework}, 
we introduced the functions 
$\bar{\bv\xi}(\theta,\bv{X},t)$ and $\bar u(\theta,\bv{X})$ 
corresponding to $\bv\xi(\bv{x},t)$ and $u(\bv{x})$,
respectively. In this Appendix, we present a method to construct
the function of $(\theta,\bv{X})$. For simplicity, we consider
functions defined in a one-dimensional interval, but the argument
presented below can be extended to functions in two- or 
higher-dimensional regions. 

Concretely, we construct a function $\bar \phi(\theta,X)$ 
corresponding to a function $\phi(x)$, where 
$\phi(x)$ is defined in the interval $[0, L]$ and 
$\bar \phi(\theta,X)$ is defined in the region $[0,\ell) \times  [0,\ell]$.
We assume that there exists an integer $N_1$ satisfying $L=N_1\ell$. 
We explain a numerical computation method to obtain the 
function $\bar \phi(\theta,X)$ from $\phi(x)$ without a rigorous 
mathematical argument.

We first divide  the interval $[0, L]$ into  small segments
$[ia,(i+1)a)$, where $0 \le  i \le  N_0N_1-1$ and $a=\ell/N_0$. 
For the function $\phi(x)$, we set 
\begin{equation}
\phi_i\equiv \phi(ia).
\end{equation}
Then, $\phi_i$ is regarded as a real-valued function from 
integers $0 \le  i \le N_0N_1-1$.
For each $i$, there exists a unique pair of the integers $i_0$ and $i_1$ 
that satisfy 
\begin{equation}
i=i_1 N_0+i_0,
\label{i0i1}
\end{equation}
where $0 \le  i_0  \le N_0-1 $ and $0 \le  i_1  \le N_1-1 $.  
Hereinafter, $i_0$ and $i_1$ are regarded as functions of $i$. 
Using this notation, we define the  function $\bar \phi_{i_0, i_1}$ as
\begin{equation}
\bar \phi_{i_0, i_1}=\phi_i.
\end{equation}

We expect that the function $\phi(x)$ appearing in physics 
is well approximated by using $\phi_i$ with a sufficiently 
small $a$. Then, for the function $\phi(x)$, we define 
$\bar\phi(\theta,X)$ by the relation
\begin{equation}
\bar \phi(\theta,X)=\bar \phi_{i_0, i_1}
\end{equation}
with $\lfloor \theta/a\rfloor=i_0$ 
and $\lfloor XL/\ell^2-\theta/\ell\rfloor=i_1$. 
Here $\lfloor x\rfloor$ is the Gauss notation that represents 
the maximum integer less than $x$. (Mathematically speaking, 
we should argue the limit $a \to 0$ and a class of functions
carefully, but this argument is not considered in this study)

Here, neglecting the irregularity arising from the 
Gauss notation, we write conventionally 
\begin{equation}
X \simeq \frac{\ell}{L}(\ell i_1+a i_0).
\end{equation}
Using Eq. (\ref{i0i1}), this implies $X \simeq \epsilon x$.
That is, the coordinate $X$ thus defined is the large-scale coordinate
describing the long distance behavior.
Next, we explain differentiation, integration, and noise 
for the functions of $(\theta,X)$.

\noindent
{\it differentiation:}
Let $\phi(x)$ be a smooth function. 
The differentiation of $\phi(x)$ is approximated by
$(\phi_{i+1}-\phi_i)/a$.
We then have
\begin{equation}
\phi_{i+1}-\phi_{i}= \bar \phi_{i_0+1,i_1}-\bar \phi_{i_0,i_1}
\end{equation}
for $0 \le i_0 \le N_0-2$. From this, we derive
\begin{align}
\phi(x+a)-\phi(x) &\simeq
\bar \phi\left(\theta+a,X+\frac{\ell}{L}a\right)-\bar \phi(\theta,X) 
\nonumber \\
 &\simeq  \partial_\theta \bar \phi(\theta,X) a
+\frac{\ell}{L}a \partial_X \bar \phi(\theta,X) +O(a^2),
\label{multi:differentiation}
\end{align}
where the approximation in the first line originates from the 
discretization error
and the fact that the irregularity of the 
Gauss notation was not considered.
Further, in the second line, we have treated $\bar \phi(\theta,X)$
as a differentiable function that might be allowed 
in the appropriate
limit $a \to 0$. From Eq. (\ref{multi:differentiation}), we obtain
\begin{equation}
\partial_x \phi(x)= (\partial_\theta +\epsilon \partial_X)
\bar \phi(\theta,X).
\end{equation}

\noindent
{\it integration:}
Let $\phi(x)$ be an integrable function. 
Then, we calculate
\begin{align}
\int {\mathrm d}x\phi(x)
\simeq &a \sum_{i=0}^{N_0N_1-1}\phi_i\nonumber\\
=& a \sum_{i_0=0}^{N_0-1}\sum_{i_1=0}^{N_1-1} \bar \phi_{i_0,i_1}
 \nonumber \\
\simeq & a \sum_{i_0=0}^{N_0-1}\sum_{i_1=0}^{N_1-1} 
\bar \phi\left(i_0 a,\frac{\ell}{L}(\ell i_1+a i_0)\right)\nonumber \\
 \simeq & \int_0^\ell \frac{d\theta}{\ell} 
\int_{0}^{\ell} \frac{{\mathrm d}X}{\epsilon} \bar \phi(\theta,X),
\label{multi:integration}
\end{align}
where the irregularity originating from the Gauss notation is
not considered (the third line) 
and the limit $a \to 0$ and $\epsilon \to 0$ is taken
(the forth line).

\noindent
{\it coordinate dependent noise:} 
Let $\xi(x,t)$ be Gaussian white noise satisfying
\begin{equation}
\bra \xi(x,t)\xi(x',t') \ket =2\delta(x-x')\delta(t-t').
\end{equation}
We set 
\begin{equation}
\phi_i(t)= \frac{1}{a}\int_{ai}^{a(i+1)}{\mathrm d}x \xi(x,t)
\end{equation}
with small $a$.  For this discretized noise, 
we define $\bar\phi_{i_0,i_1}(t)$ and $\bar \phi(\theta,X,t)$
in the same manner as that for the case in which $\phi_i=\phi(ia)$.
Conventionally, we denote $\bar \phi(\theta,X,t)$ as $\bar\xi(\theta,X,t)$
(see \ref{basicframework}).
Based on the definitions described above, we can derive
\begin{align}
&\bra\bar\xi(\theta,X,t)\bar\xi(\theta',X',t') \ket\nonumber\\
&=2\epsilon \ell  \delta(\theta-\theta')\delta(X-X')\delta(t-t').
\label{noiseintensityinmulti}
\end{align}

Let $\varphi(\theta)$ be a smooth function that satisfies 
$\varphi(0)=\varphi(\ell)$. We denote 
the Stratonovich product of $\varphi(\theta)$ and $\bar\xi(\theta,X,t)$ 
as $\varphi(\theta) \circ \bar \xi(\theta,X,t)$. 
This product can be written by
using the discretized form $\phi_{i_0,i_1}$
with an additional definition 
$\bar \phi_{N_0,i_1}=\bar \phi_{0,i_1}$. From this, we obtain
\begin{equation}
\int_0^\ell {\mathrm d}\theta 
\partial_\theta[ \varphi(\theta) \circ\bar\xi(\theta,X,t)]=0.
\end{equation}
This formula is used for
obtaining Eq. (\ref{omega1non}).

\section{proof of Eqs. (\ref{r11})-(\ref{B11D})}\label{D=D}

In this Appendix, we present the proofs of 
Eqs. (\ref{r11}), (\ref{B22D}), (\ref{Df}), and (\ref{B11D}).
We first prove the key equality
\begin{align}
\frac{ d}{d \theta}\left(\frac{b(\theta)}{\ps(\theta)}\right)+1=&
\left(\int_0^\ell\frac{{\mathrm d}\theta'}{\ell}I_-(\theta')\right)^{-1}
 I_+(\theta).
\label{partialb3}
\end{align}
All the equations can be derived from Eq. (\ref{partialb3}).

\subsection{proof of the key equality}

We first derive an explicit form of $b(\theta)$ defined in Eq. (\ref{bdef}). 
Applying $\hat{M}^{(0)\dagger}$ given by Eq. (\ref{defofMdagger})
to both the left- and right-hand sides 
of Eq. (\ref{bdef}), we obtain the following differential equation 
for $b(\theta)$:
\begin{align}
&\frac{ d }{d\theta}\left(
\frac{\Js}{\ps(\theta)}b(\theta)
-\frac{T}{\gamma}\der{ \log \ps(\theta)}{\theta}b(\theta)
+\frac{T}{\gamma}\der{b(\theta)}{\theta}\right)\nonumber\\
&=\Js(\ps(\theta)-1)-\frac{T}{\gamma}\der{\ps(\theta)}{\theta}.
\label{b:differentialeq2}
\end{align}
Integrating Eq. (\ref{b:differentialeq2}), we obtain the following 
first-order 
differential equation for  $b$:
\begin{align}
\frac{\Js}{\ps}b(\theta)&
-\frac{T}{\gamma}\der{ \log\ps(\theta) }{\theta}b(\theta)
+\frac{T}{\gamma}\der{b(\theta)}{\theta}\nonumber\\
&=\Js\left[H(\theta)-\theta-K_1\right]-\frac{T}{\gamma}\ps(\theta),
\label{b:differentialeq1}
\end{align}
where $K_1$ is a constant whose value is determined later.
Here, $H(\theta)$ and $V(\theta)$ are defined as
\begin{align}
H(\theta)\equiv&\int_0^\theta {\mathrm d}\theta' \ps(\theta'),\\
V(\theta)\equiv&U(\theta)-f\theta.
\end{align}

We introduce $\bar b$ in the equation
\begin{align}
b(\theta)=\ps(\theta)\left[H(\theta)-\theta+\bar b(\theta)\right],
\label{defofbarb}
\end{align}
and then rewrite  Eq. (\ref{b:differentialeq1}) as
\begin{align}
\Js \left[
\bar b(\theta)+K_1\right]
 +\frac{T}{\gamma}\ps(\theta)
 \left(\ps(\theta)+\der{\bar b(\theta)}{\theta}\right)=0.
\label{barb:differentialeq1}
\end{align}
Using the equality
\begin{align}
\Js \ps \e^{\beta V}=-\frac{T}{\gamma}\ps
\frac{ d }{d\theta}\left(\ps\e^{\beta V}
\right),
\label{jspsV}
\end{align}
we obtain the solution of Eq. (\ref{barb:differentialeq1}) as
\begin{align}
\bar b(\theta)=-K_1 +\ps(\theta)\e^{\beta V(\theta)}
[K_2-G(\theta)],
\label{barb}
\end{align}
where $K_2$ is a constant that is determined later, and $G(\theta)$ is defined as
\begin{align}
G(\theta)=&\int_0^\theta{\mathrm d}\theta'
\e^{-\beta V(\theta')}.
\end{align}
Substituting Eq. (\ref{barb}) into Eq. (\ref{defofbarb}), 
we write
\begin{align}
\frac{b(\theta)}{\ps(\theta)}=
H(\theta)-\theta-K_1
+\ps(\theta)\e^{\beta V(\theta)}
[K_2-G(\theta)].
\label{result:b}
\end{align}
Now, $K_1$ and $K_2$ are determined from the conditions
$(b,\Phi_0)=0$ and $b(0)=b(\ell)$.
The results are as follows:
\begin{align}
K_2=&\frac{1}{1-\e^{\beta f \ell}}G(\ell),
\label{K4}\\
K_1=&\int_0^\ell \frac{{\mathrm d}\theta}{\ell}\left\{
\ps(\theta)[H(\theta)-\theta]
+\ps^2(\theta)\e^{\beta V(\theta)}[K_2-G(\theta)]
\right\}.
\label{K3}
\end{align}

Next, we note the identity
\begin{align}
&\int_0^\ell{\mathrm d}\theta'\phi(\theta')\e^{\beta f\theta'}
-(1-\e^{\beta f\ell})
\int_0^\theta{\mathrm d}\theta'\phi(\theta')\e^{\beta f\theta'}\nonumber\\
&=\e^{\beta f\theta}\int_0^\ell{\mathrm d}\theta'\phi(\theta'+\theta)
\e^{\beta f\theta'}
\label{identity}
\end{align}
for an arbitrary periodic function $\phi(\theta)$ with period $\ell$.
Setting $\phi=\e^{-\beta U}$ in Eq. (\ref{identity}), we obtain
\begin{align}
K_2-G(\theta)=\frac{1}{1-\e^{\beta f\ell}}\e^{-\beta V(\theta)}I_+(\theta).
\label{simple:b0}
\end{align}
Then, substituting Eq. (\ref{simple:b0}) into Eq. (\ref{result:b})
and multiplying $\Js$ to both the left- and right-hand sides, 
we obtain
\begin{align}
\Js\frac{b(\theta)}{\ps(\theta)}
=&\Js[H(\theta)-\theta-K_1]\nonumber\\
&-\frac{T}{\gamma}
\left(\int_0^\ell \frac{{\mathrm d}\theta'}{\ell}I_-(\theta')\right)^{-1}
\ps(\theta)I_+(\theta),
\label{simple:b}
\end{align}
where we have used Eq. (\ref{defofJs}).

On the other hand, 
we rewrite Eq. (\ref{b:differentialeq1}) as
\begin{align}
\frac{T}{\gamma}&\ps(\theta)\left[\frac{ d }{d\theta}
\left(\frac{b(\theta)}{\ps(\theta)}\right)+1\right]\nonumber\\
&=-\frac{\Js}{\ps}b+\Js(H(\theta)-\theta-K_1).
\label{b:differentialeq1'}
\end{align}
Comparing Eqs. (\ref{simple:b}) and (\ref{b:differentialeq1'}),
we obtain Eq. (\ref{partialb3}).

\subsection{proof of Eq. (\ref{Df})}

We can rewrite $D$ in Eq. (\ref{result:D}) as
\begin{align}
D=&\Js\int_0^\ell \frac{{\mathrm d}\theta}{\ell}
\frac{b(\theta)}{\ps(\theta)}
+\frac{T}{\gamma}\int_0^\ell \frac{{\mathrm d}\theta}{\ell}
\left[
\frac{ d }{d\theta}
\left(\frac{b(\theta)}{\ps(\theta)}\right)+1\right]
\ps(\theta).
\label{anotherD}
\end{align}
Then, substituting Eqs. (\ref{simple:b}) and (\ref{partialb3}) 
into Eq. (\ref{anotherD}), we obtain
\begin{align}
D=&\Js\int_0^\ell \frac{{\mathrm d}\theta}{\ell}
[H(\theta)-\theta]-\Js K_1.
\label{anotherD2}
\end{align}
Substituting Eq. (\ref{K3}) into Eq. (\ref{anotherD2}) and 
using Eq. (\ref{simple:b0}), we rewrite (\ref{anotherD2}) as
\begin{align}
D=-\Js\int_0^\ell \frac{{\mathrm d}\theta}{\ell}
\ps^2(\theta)\e^{\beta V(\theta)}\left(
\frac{1}{\e^{\beta f\ell}-1}\e^{-\beta V(\theta)}I_+(\theta)\right),
\label{anotherD3}
\end{align}
where we have used the identity
\begin{align}
\int_0^\ell\frac{{\mathrm d}\theta}{\ell}[\ps(\theta)-1][\theta-H(\theta)]=0.
\end{align}
Finally, substituting Eqs. (\ref{defofJs}) and (\ref{Ipm:def}) 
into Eq. (\ref{anotherD3}), we obtain Eq. (\ref{Df}).

\subsection{proof of Eqs. (\ref{r11}), (\ref{B22D}), and (\ref{B11D})}
We first calculate the correlation function of $\eta_2(\bv{X},t)$
as follows:
\begin{widetext}
\begin{align}
\langle\eta_2(\bv{X},t)\eta_2(\bv{X}',t')\rangle
&=\frac{T}{\gamma}
\int_0^\ell\frac{{\mathrm d}\theta}{\ell}
\int_0^\ell\frac{{\mathrm d}\theta'}{\ell}
\frac{\Psi_0(\theta)}{\sqrt{\ps(\theta)}}
\frac{\Psi_0(\theta')}{\sqrt{\ps(\theta')}}
\bra\bar\xi_2(\theta,\bv{X},t)\bar\xi_2(\theta',\bv{X}',t')\ket
\nonumber\\
&=\frac{T}{\gamma}
\int_0^\ell\frac{{\mathrm d}\theta}{\ell}
\int_0^\ell\frac{{\mathrm d}\theta'}{\ell}
\frac{\Psi_0(\theta)\Psi_0(\theta')}{\sqrt{\ps(\theta)\ps(\theta')}}
2\ell\epsilon^2\delta^2(\bv{X}-\bv{X}')
\delta(\theta-\theta')\delta(t-t')\nonumber\\
&=\frac{2T}{\gamma}\epsilon^2\delta^2(\bv{X}-\bv{X}')\delta(t-t'),
\label{result:B22D}
\end{align}
\end{widetext}
where we have used (\ref{noiseintensityinmulti}).
Using Eqs. (\ref{coarse3}) and (\ref{devnon}), we obtain Eq. (\ref{B22D}).

Next, we consider the correlation function of 
$\zeta(\bv{X},t)+\eta_1(\bv{X},t)$ as follows:
\begin{widetext}
\begin{align}
\langle&[\zeta(\bv{X},t)+\eta_1(\bv{X},t)]
[\zeta(\bv{X}',t')+\eta_1(\bv{X}',t')]\rangle\nonumber\\
&=\frac{T}{\gamma}
\int_0^\ell\frac{{\mathrm d}\theta}{\ell}
\int_0^\ell\frac{{\mathrm d}\theta'}{\ell}
\left[
\frac{ d }{d\theta}\left(\frac{b(\theta)}{\ps(\theta)}\right)
+1\right]
\left[
\frac{ d }{d\theta'}\left(\frac{b(\theta')}{\ps(\theta')}\right)
+1\right]\sqrt{\ps(\theta)\ps(\theta')}
\bra\bar\xi_1(\theta,\bv{X},t)\bar\xi_1(\theta',\bv{X}',t')\ket\nonumber\\
&=\frac{2T}{\gamma}
\int_0^\ell\frac{{\mathrm d}\theta}{\ell}
\left[\frac{d}{d\theta}\left(\frac{b(\theta)}{\ps(\theta)}\right)
+1\right]^2\ps(\theta)\epsilon^2\delta^2(\bv{X}-\bv{X}')\delta(t-t').
\label{rep:B11}
\end{align}
\end{widetext}
This corresponds to Eq. (\ref{r11}).

Finally,
substituting Eqs. (\ref{partialb3}) and (\ref{defofJs}) 
into Eq. (\ref{r11}), 
we obtain 
\begin{align}
B_{11}
=&\frac{T}{\gamma}
\left(\int_0^\ell\frac{{\mathrm d}\theta}{\ell}I_-(\theta)\right)^{-3}
\int_0^\ell\frac{{\mathrm d}\theta}{\ell}I_-(\theta)
\left[I_+(\theta)\right]^2.
\label{noisefinal}
\end{align}
Because the right-hand side of Eq. (\ref{noisefinal}) is invariant
for the exchange of $I_+$ and $I_-$ (see Ref. \cite{Reimann2}), 
$B_{11}$ in Eq. (\ref{noisefinal}) is equal to $D$ in Eq. (\ref{Df}). 
This corresponds to Eq. (\ref{B11D}).

\end{document}